\documentclass[pra,amsmath,amssymb,twocolumn,longbibliography]{revtex4-1}
\usepackage{graphicx}
\usepackage{dcolumn}
\usepackage{bm}
\usepackage[utf8]{inputenc}
\usepackage[english]{babel}
\usepackage{amsfonts}
\usepackage{hyperref}
\usepackage{physics}
\usepackage{epstopdf}
\usepackage{mathtools}
\usepackage[capitalise]{cleveref}
\usepackage{float}
\usepackage[space]{grffile}
\usepackage{color}
\usepackage[normalem]{ulem}
\usepackage{subfigure}
\usepackage{natbib}

\binoppenalty=\maxdimen
\relpenalty=\maxdimen

\addto\captionsenglish{}

\def \htx{\hat{\tilde{x}}}
\def \htp{\hat{\tilde{p}}}
\def \hta{\hat{\tilde{a}}}

\def \G0{\tilde{G}_0^R}

\def \hx{\hat{x}}

\def \hp{\hat{p}}

\def \hX{\hat{X}}
\def \hP{\hat{P}}

\def \hB{\hat{B}}

\def \ha{\hat{a}}

\def \hH{\hat{H}}

\def \hM{\hat{M}}
\def \hfancyB{\hat{\mathcal{B}}}

\begin{document}
	
	\title{Exponentially-enhanced quantum sensing with non-Hermitian lattice dynamics}
	
	\author{A. McDonald$^{1,2}$ and A. A. Clerk$^1$}
	\affiliation{$^1$Pritzker School of Molecular Engineering, University of Chicago, Chicago, IL 60637, USA\\
		$^2$Department of Physics,
		University of Chicago, Chicago, IL 60637, USA}
	

	\begin{abstract}
    We study how unique features of non-Hermitian lattice systems can be harnessed to improve Hamiltonian parameter estimation in a fully quantum setting.  While the so-called non-Hermitian skin effect does not provide any distinct advantage, alternate effects yield dramatic enhancements.  We show that certain asymmetric non-Hermitian tight-binding models with a $\mathbb{Z}_2$ symmetry yield a pronounced sensing advantage: the quantum Fisher information per photon increases exponentially with system size.  We find that these advantages persist in regimes where non-Markovian and non-perturbative effects become important.  Our setup is directly compatible with a variety of quantum optical and superconducting circuit platforms, and already yields strong enhancements with as few as three lattice sites.
	\end{abstract}

	\maketitle
	
	\section{Introduction}

	Quantum metrology and sensing aim to improve measurement precision over classical devices by exploiting uniquely quantum phenomena such as entanglement and squeezing \cite{Giovannetti2011,DegenRMP,RMP_Spins}.  It is interesting to ask whether distinct
	effects associated with non-Hermitian dynamics can also be used to improve sensors operating in quantum regimes \cite{Langbein_2018, Kero_Nat_Comm,Liang_2019,Liu2019,Murch_2019}.  In purely classical settings, mode degeneracies specific to non-Hermitian systems (so-called exceptional points) have been suggested as a means for enhanced parametric sensing \cite{Wiersig_2014}.  Evidence for enhancement has been demonstrated in several classical-domain experiments involving small coupled mode systems  (see e.g.~Refs.~\onlinecite{Lan_Yang_2017, Christodoulides_2017, Alu_2019, Vahala_2019}).  Theory suggests that particular kinds of non-Hermitian effects could also be useful in truly quantum settings \cite{Kero_Nat_Comm}.
	
    To date, both theory and experiment have focused on non-Hermitian sensing schemes that utilize at most a few coupled modes.  It is however well known that unusual new phenomena appear when considering genuinely multi-mode non-Hermitian dynamics. The paradigmatic example is the so-called ``non-Hermitian skin effect" \cite{Zhong_PRL_2018_1, Lee_2016, Alexander_2018}, which occurs in several non-Hermitian tight-binding models \cite{Xiong_2017, Thomale_2019, Udea_PRX_2019}. In these systems, all eigenvalues and wavefunctions of the Hamiltonian exhibit a dramatic sensitivity to a change of boundary conditions.  This extreme sensitivity would seem to be a potentially powerful resource for parametric sensing \cite{Schomerus_2020}.
	
	\begin{figure}[t]
		\centering
		\includegraphics[width=0.45\textwidth]{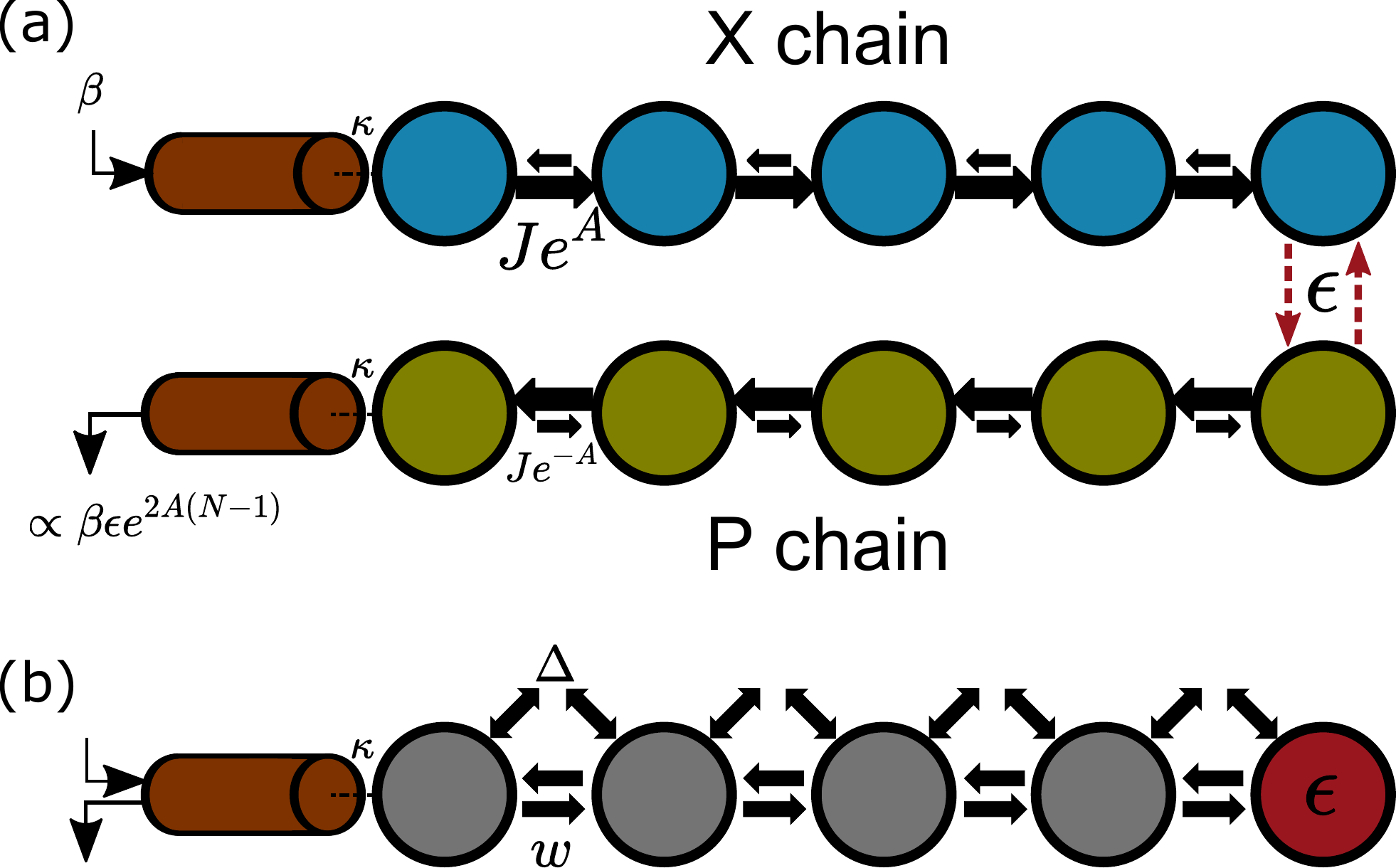}
		\caption{
		(a) Basic lattice sensor: two $N$-site 
		non-Hermitian tight binding chains, each with opposite chirality.  Each chain has asymmetric hopping:  for the top (bottom) chain, hopping to the right is a factor of $e^{2A}$ larger (smaller) than hopping to the left.  The two lattices are only coupled via a weak symmetry breaking perturbation $\epsilon$ on the rightmost site; the goal is to estimate $\epsilon$.  A signal entering the top X chain induces an exponentially large output in the bottom P chain, but only if $\epsilon \neq 0$. (b) An array of bosonic cavities coupled via nearest neighbour hopping $w$ and coherent two-photon drive $\Delta$ with a small detuning $\epsilon$ on the last site. This provides a dissipation-free realization of the setup in (a), where the canonical quadratures $\hx$ and $\hp$ play the role of the top and bottom chains respectively.  This system yields an exponentially enhanced SNR even when quantum noise effects are included. }
		\label{fig:Schematic}
	\end{figure}

	In this work, we show that non-Hermitian lattice dynamics does indeed provide a unique means for constructing enhanced sensors; moreover, this advantage persists even when operating in truly quantum regimes.  We study in detail Hamiltonian parameter estimation using a one-dimensional lattice model with asymmetric tunneling (akin to the well-studied Hatano-Nelson model \cite{Hatano_Nelson}).  We find, somewhat surprisingly, that the non-Hermitian skin effect does not provide any advantage over more traditional sensing protocols. Rather, we find another distinct non-Hermitian mechanism that enables a dramatic enhancement of measurement sensitivity:  the quantum Fisher information per photon exhibits an exponential scaling with system size.  As we discuss, the underlying mechanism makes use of both non-reciprocity and an unusual kind of symmetry breaking.
	
	While our ideas are general, our analysis focuses on a system that uses parametric driving to realize non-Hermitian dynamics; this has the strong advantage of not requiring any external dissipation or post-selection \cite{Alexander_2018,Yuxin_2019}. Further, we ultimately focus on dispersive sensing, where the parameter of interest shifts the frequency of a resonant mode. This is a ubiquitous sensing strategy, with applications ranging from superconducting qubit measurement \cite{Circuit_QED_PRA} to virus detection \cite{Vollmer2008}. Our proposal is also compatible with a number of different experimental platforms in superconducting quantum circuits and quantum optics, and ultimately requires one to make a standard homodyne measurement.  We also consider physics that goes beyond the usual limit of strictly infinitesimal parameter sensing.  We find that the exponential enhancement of measurement sensitivity persists even when considering limitations associated with the finite propagation time of a large lattice.  Even for parameters large enough to invalidate a full linear response analysis, we find that our scheme provides a strong advantage:  it achieves a square-root enhancement of the sensitivity (including noise effects).  This is similar to what is found in exceptional point sensors in the absence of noise \cite{Wiersig_2014}.  Finally, while our discussion focuses on large lattices, the results we present are already interesting in a small system consisting of just three coupled resonators.

	\section{Ingredients for a non-Hermitian lattice sensor}

    \subsection{Amplified non-reciprocal response in the Hatano-Nelson model}

    A key feature that we will exploit in our new sensor is the dramatically large and uni-directional response exhibited by certain non-Hermitian lattice models:  perturbing a single lattice site induces a large change at one end of the chain, but not the other (see e.g.~\cite{Schomerus_2020, Nunnenkamp_2019}).  We start by providing a physically-transparent explanation of this effect, based on interpreting non-Hermitian asymmetry in tight-binding matrix elements as directional gain and loss.
    
    The simplest relevant system is the well-known Hatano-Nelson model \cite{Hatano_Nelson, Hatano_Nelson_2}.  This is a 1D tight-binding chain with asymmetric nearest-neighbour hoppings, 
    $\hat{H} = i J \sum_n \left( e^A \ketbra{n+1}{n} - e^{-A} \ketbra{n}{n+1} \right) $, where $J, A$ are real and $\ket{n}$ is a position eigenket. 
    The corresponding single-particle Schr\"{o}dinger equation is ($\hbar = 1$ throughout)
	\begin{align}\label{eq:Hatano-Nelson}
    	\dot{\psi}_n = J e^{A}  \psi_{n-1} - J e^{-A} \psi_{n+1}, 
	\end{align}
	where $\psi_n = \braket{n}{\psi}$.
    While $A$ formally plays the role of an imaginary vector potential, it is more usefully thought of as an amplification factor. Assuming $A$ is positive for definiteness, Eq.~(\ref{eq:Hatano-Nelson}) describes a system where a wavefunction's amplitude grows by $e^{A}$ every time a particle hops one site to the right, and decays an equal amount $e^{-A}$ as it travels to the left, regardless of its energy.
	
	With this picture in mind, the form of the real-space susceptibility (i.e.~single particle Green's function)  $\chi(n,m;t)$ for a finite open chain has an intuitive form.
	Letting $\ket{m(t)} = e^{-i \hat{H} t} \ket{m}$, a simple calculation yields (see App.~\ref{app:chi_quadrature}):
	\begin{align} 
	    \chi(n,m;t)
	    & \equiv
	        \braket{n}{m(t)}
    	\label{eq:Susceptibility}
    	=
    	e^{A(n-m)}
    	\chi_0(n,m;t).
	\end{align}
 	Here, $\chi_0(n,m;t)$ is the susceptibility matrix when $A=0$, 
 	i.e.~the Green's function of a Hermitian tight-binding chain.  This quantity is reciprocal, in the sense that  $\chi_0(n,m;t) = (-1)^{m-n}\chi_0(m,n;t)$ (i.e.~apart from a phase, there is no asymmetry in rightwards versus leftwards propagation).  The Green's function $\chi_0(n,m;t)$ both describes how particles propagate in the lattice, and also the response properties of the system (i.e.~if you perturb site $m$ at $t=0$, how does site $n$ respond at some later time?).  
 	
 	The simple factorization in Eq.~(\ref{eq:Susceptibility}) makes it clear that there are two basic processes determining the response.  The first is a distance and direction-dependent amplification / deamplification factor, whereas the second encodes the dynamics of the underlying ($A=0$) Hermitian tight-binding model. 
 	We thus have a simple intuitive picture for the susceptibility, without having to make recourse to other seemingly more complicated non-Hermitian features, such as exceptional points, the non-Hermitian skin effect, or the Petermann factor \cite{Petermann_1979, Grangier_1998}. 
 	Note that Eq.~(\ref{eq:Susceptibility}) can be easily derived via a similarity transformation, which is analogous to the gauge transformation one would make if $A$ were imaginary (and hence a real synthetic gauge field) \cite{Hatano_Nelson, Hatano_Nelson_2}.  
 	
 	\subsection{$\mathbb{Z}_2$ symmetry in non-Hermitian lattice models}
 	
 	The second basic ingredient we will exploit in constructing our sensor is  symmetry breaking.  The Hatano-Nelson chain breaks reciprocity for any $A \neq 0$; formally, it picks a preferred amplification direction, and does not remain invariant (up to a local gauge change) under a spatial inversion operation $\ket{n} \rightarrow \ket{-n}$.  We can trivially restore this symmetry by considering a system with {\it two} uncoupled Hatano-Nelson chains indexed by $\sigma = \uparrow, \downarrow$ with amplification factors $A_{\uparrow}, A_{\downarrow}$.  If we pick $A_{\uparrow} = -A_{\downarrow}$, then the composite system restores some of the lost symmetry.  Formally, the two-chain system is invariant up to a local gauge change under the combined operations $\ket{n} \rightarrow \ket{-n}$ (spatial inversion) and $\sigma \rightarrow \bar{\sigma} $ (pseudospin inversion).  While this may seem trivial, this kind of discrete symmetry can persist even for certain forms of interchain coupling, and has recently been interpreted as a formal $\mathbb{Z}_2$ symmetry class with its own distinct non-Hermitian topological phenomena \cite{Sato2020}. We discuss this symmetry more formally in Appendix \ref{app:Symmetry}.
 	
 	  For our purposes, the interesting feature here will be to consider {\it breaking} this symmetry with an external perturbation whose magnitude we wish to estimate.  As we will see, the response to this symmetry breaking
 	  can be exponentially large in system size, enabling a new kind of sensor.


	\section{Model and Measurement Protocol}

	With the motivation of the previous section, we now consider a sensor comprised of two Hatano-Nelson chains with an opposite chirality 
	(see Fig.~\ref{fig:Schematic}(a)).  There are a variety of means for such realizing non-Hermitian directional tight-binding models using dissipation  \cite{Metelmann2015,Metelmann2018}; approaches based on feedback control are also possible and have been recently implemented \cite{Coulais2019}.
	However, for optimal sensing properties in quantum settings, methods that are both autonomous and avoid the noise associated with dissipation are desirable.  We thus focus on a dissipation-free method for realizing non-Hermitian dynamics based on parametric driving \cite{Alexander_2018,Yuxin_2019}.  We stress that the response properties of our sensor will be independent of how the non-Hermitian dynamics is implemented, and hence apply equally well to dissipative and feedback based strategies. 
	
	We consider an $N$-site chain of driven, coupled bosonic modes described by the  
    fully Hermitian Hamiltonian 
	\begin{align}
	\label{eq:HBKC}
	\hH_B =  
	\sum_{n =1 }^{N-1}
	\left( i w  \ha^{\dagger}_{n+1} \ha_n + 
	i \Delta  \ha^{\dagger}_{n+1} \ha_{n}^\dagger + h.c. \right).
	\end{align}
	Here $\hat{a}_j$ is the photon annihilation operator on site $j$, $w$ is the nearest-neighbour hopping term, $\Delta$ is the nearest-neighbour two-photon drive, and we consider open boundary conditions. We take both $w$ and $\Delta$ to be positive and  $w > \Delta$. This model describes a 1D cavity array subject to parametric drives on each bond (described in a rotating frame set by the external pump frequency).
	As discussed extensively in 
	Ref.~\onlinecite{Alexander_2018}, this system could be realized in both quantum superconducting circuits or nonlinear quantum optical systems.  Note the lack of any on-site terms corresponds to the parametric driving frequency matching the resonance frequency of each isolated cavity.

	Although not immediately obvious, the dynamics generated by $\hH_B$ 
	corresponds to two copies of the Hatano-Nelson model. In the basis of local canonical quadrature operators $\hx_j$ and $\hp_j$, defined via $\ha_j = (\hx_j+i\hp_j)/\sqrt{2}$, the Hamiltonian reads
	\begin{align}\label{eq:H_B_xp}
	   \hH_B
	   =
	   \sum_{n=1}^{N-1}
    \left(
	-(w-\Delta) \hx_{n+1} \hp_n + (w+\Delta) \hp_{n+1} \hx_n
	\right).
	\end{align}
	This then yields the Heisenberg equations of motion
	\begin{align}
	\label{eq:X_EOM}
	\dot{\hx}_n & = 
	J e^{A} \hx_{n-1} - 
	J e^{-A} \hx_{n+1},\\
	\label{eq:P_EOM}
	\dot{\hp}_n & = 
	J e^{-A} \hp_{n-1} - 
	J e^{A} \hp_{n+1},
	\end{align}
    where the effective hopping amplitude $J$ and imaginary vector potential $A$ are related to $w$ and $\Delta$ by
	\begin{align}
	\label{eq:DefJ}
	&J = \sqrt{w^2-\Delta^2},
	\\
	\label{eq:DefA}
	&e^{2A}
	=
	\frac{w+\Delta}{w-\Delta}.
	\end{align}
	Comparing against Eq.~(\ref{eq:Hatano-Nelson}), we see that the dynamics of each canonical quadrature corresponds to that of a Hatano Nelson model, with opposite chiralities for $\hat{x}$ and $\hat{p}$ (Fig.~\ref{fig:Schematic}).  These orthogonal quadratures correspond to different phases of photonic excitations, and hence the system exhibits phase-dependent non-reciprocal amplification \cite{Alexander_2018}. Note that there is a constraint on our mapping:  the complex wavefunction amplitudes in the Hatano-Nelson model have been replaced by Hermitian quadrature operators in our system.  This will play no role in what follows.  
	
	We now demonstrate how this setup can be used for Hamiltonian parameter estimation.  We add a Hermitian perturbation $\epsilon \hat{V}$ to our  Hamiltonian where $\hat{V}$ is some system operator; the goal is to estimate $\epsilon$.  
	We also couple the first site of our lattice to an input-output waveguide as a means to probe its properties.  The simplest protocol is to use this waveguide to drive the system with a classical tone (i.e.~a coherent state), and then measure the outgoing light in the waveguide (see Fig.~\ref{fig:Schematic}(b)). 
	The full Hamiltonian becomes
	\begin{align}
	    \hH[\epsilon]
	    =
	    \hH_B
	    +
	    \epsilon \hat{V}
	    +
	    \hH_\kappa
	    -i\sqrt{\kappa}
	    \left(
	    \ha^\dagger_1 \beta
	    -
	    h.c.
	    \right)
	\end{align}
	 $\hH_\kappa$ describes damping of the first site at a rate $\kappa$, due to coupling to the modes of the waveguide which we treat using standard input-output theory \cite{RMP_Clerk}.
	 The last term corresponds to a classical drive with amplitude $\beta = |\beta| e^{i \theta}$. Note that we take the drive frequency to match the resonance frequency of the isolated cavities; this frequency is zero in our rotating frame.  
	 
	 Using the standard input-output boundary condition, the output field in the waveguide is given by
	\begin{align}\label{eq:In_Out}
	    \hB^{\rm (out)}(t)
	    =
	    \left(
	    \beta + \hB^{\rm (in)}(t)
	    \right)
	    +\sqrt{\kappa} \ha_1(t)
	\end{align}
	where $\hB^{\rm (in)}$, the operator equivalent of Gaussian white noise, describes the noise entering the lattice through the waveguide. Our goal is to estimate $\epsilon$ by making an optimal measurement of the output field.  
	In what follows, we take $\epsilon$ to have units of frequency and $\hat{V}$ to be dimensionless.

    We further specialize to the usual case where $\epsilon$ is so small that it can only be estimated by integrating the output field over a long timescale $\tau$. If we turn on the drive tone at $t=0$, the relevant temporal mode of the output field to consider is
	\begin{align}
	    \hfancyB_\tau(N)
	    =
	    \frac{1}{\sqrt{\tau}}
	    \int_0^{\tau}
	    dt
	    \hB^{\rm (out)}
	    (t)
	\end{align}
	Note that this is normalized to be a canonical bosonic lowering operator, satisfying $[\hfancyB_\tau(N),\hfancyB_\tau^\dagger(N)] = 1$.
	We write an explicit dependence on the chain size $N$, as we will be interested in understanding how things scale as $N$ is increased.  	
	
	The maximum amount of information available in $\hfancyB_\tau(N)$ on $\epsilon$ is quantified by the quantum Fisher information (QFI). The QFI provides a lower bound on the root mean square error of any (unbiased) estimate of $\epsilon$ regardless of how  $\hfancyB_{\tau}(N)$ is measured \cite{Giovannetti2011}.
    Calculation of the QFI unfortunately does not in general tell one the form of the optimal measurement.  However, in our linear Gaussian system, things are much simpler:  for large $| \beta|$, the optimal measurement will always correspond to a standard homodyne measurement \cite{Pirandola_2015, Braun_2013}. 
    The relevant Hermitian measurement operator has the form
	\begin{align}\label{eq:Meas_Op}
	    \hM_\tau(N)
	    =
	   \frac{1}{\sqrt{2}}
        \left(
        e^{-i \phi}
	    \hfancyB_\tau(N)
	    +
	    e^{i \phi}
	    \hfancyB_\tau^\dagger(N)	  
	    \right),
	\end{align}
	i.e.~a quadrature of the output operator $\hfancyB_\tau(N)$ along a direction in phase space determined by the angle $\phi$.
	
	We will focus throughout on the large-drive limit, and will be interested in characterizing the QFI to leading order in $|\beta|$.  In this limit, QFI is determined by the statistics of $\hM_\tau(N)$ via  \cite{Kero_Nat_Comm, Pirandola_2015}
	\begin{align}\label{eq:QFI}
	   \mathrm{QFI}_{\tau}(N)
	    &=
	    \max_{\phi}
	    \left[
	    \lim_{\epsilon \to 0}
	    \left(
	    \frac{1}{\epsilon}
	    \frac
	    {
        \mathcal{S}_\tau(N, \epsilon)
        }
	    {
	    \mathcal{N}_{\tau}(N, \epsilon)
	    }
	    \right)^2
	    \right],
	\end{align}
	where
	\begin{align}
	    &
	    \mathcal{S}_{\tau}(N, \epsilon)
	    =
	    |
	    \langle \hM_\tau(N) \rangle_{\epsilon}
	    -
	    \langle \hM_\tau(N) \rangle_{0}
	    |,
	    \label{eq:SignalDef}
	    \\
	    &
	    \mathcal{N}_{\tau}(N, \epsilon)
	    =
	    \sqrt{
	    \langle
	    \hM^2_\tau(N)
	    \rangle_\epsilon
	    -
	    \langle
	    \hM_\tau(N)
	    \rangle^2_\epsilon,
	    }
	\end{align}
	are the signal and the noise respectively associated with the measurement. Here, $\langle \cdot \rangle_z$ means an average with respect to a state whose dynamics are governed by $\hH[z]$.

	This expression for the QFI coincides with the SNR of an optimal homodyne measurement, and scales as $| \beta |^2$; the next-leading order term is independent of $|\beta|$. Note that the QFI only depends on the noise $\mathcal{N}_{\tau}(N, \epsilon)$ calculated to zeroth order in $\epsilon$.
	We stress that the expression for the QFI still depends on the drive phase $\theta$ as well as the form of the operator $\hat{V}$; in what follows, we will be interested in optimizing these as well.  

    Given its role as a fundamental performance metric, it is tempting to declare that a better sensor has been built if it increases the QFI. Different measurement strategies however use resources differently, and one must carefully consider which to constrain when making comparisons.  In our case, we wish to distinguish a true sensing enhancement from a more trivial effect, where a different protocol simply results in there being more photons in the system available to interact with the perturbation $\hat{V}$ 
    (as occurs with standard exceptional-point based sensing schemes \cite{Kero_Nat_Comm,Liang_2019}).
    For this reason, we will take as the relevant metric the QFI scaled by the {\it total} average photon number $\bar{n}_{\rm tot}$ \cite{Kero_Nat_Comm}:
		\begin{align}
	    \bar{n}_{\rm tot} \equiv \sum_n \langle \ha_n^\dagger \ha_n \rangle_0
	    \simeq  
	    \sum_n 
	    \langle \ha_n^\dagger  \rangle_0
	    \langle \ha_n \rangle_0 \propto |\beta|^2/\kappa.
	\end{align}
	As we consider throughout the large-drive limit, we only keep the leading-order-in-$\beta$ contribution to $\bar{n}_{\rm tot}$.  This is simply the photon number associated with the drive-induced displacement of each cavity annihilation operator.  The additional contribution to 
	$\bar{n}_{\rm tot}$ due to amplification of vacuum fluctuations is $\beta$-independent, hence plays no role in the large-drive limit we consider (see Appendix \ref{app:n_bar}). 

	\begin{figure*}
		\centering
		\includegraphics[width=1\textwidth]{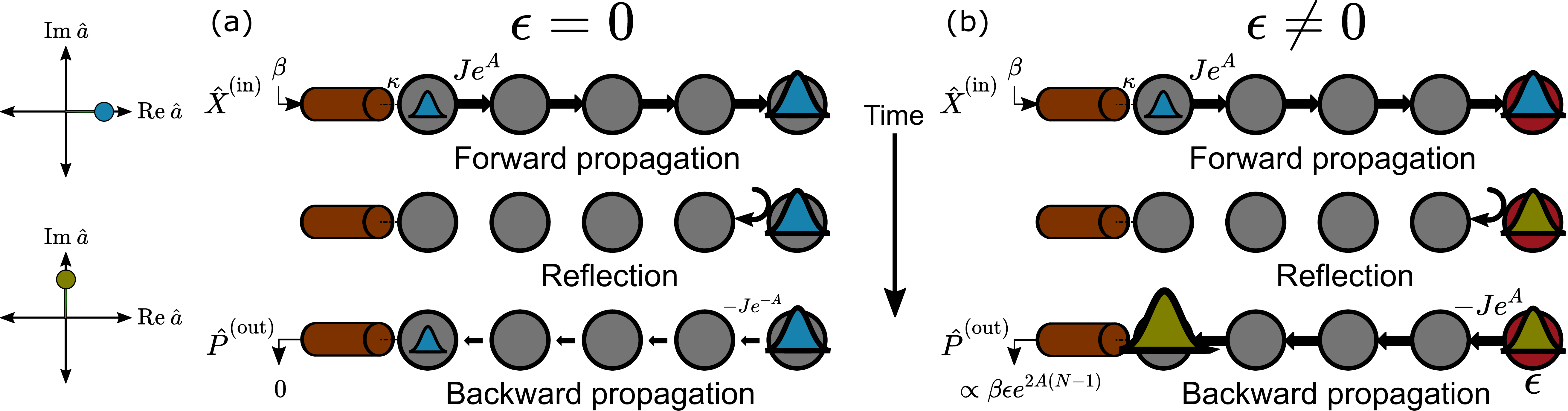}
		\caption{
		Schematic of measurement dynamics.  A classical drive is injected into the leftmost lattice site via a waveguide (coupling rate $\kappa$).  The drive amplitude is real (blue wavepacket), 
		corresponding to an $\hx$ quadrature excitation. As the wavepacket propagates rightwards, its amplitude grows $\propto e^{A(N-1)}$ until it reaches the last site $N$. (a) If $\epsilon = 0$, the wavepacket scatters off the open boundary without changing quadrature.  It is thus 
		{\it deamplified} as it propagates back to the first site.  As a result, for $\epsilon=0$ there is no amplification of the drive or of injected noise.  (b) For non-zero $\epsilon$, a wavepacket can scatter off the boundary and change quadrature (olive wavepacket).  It then is {\it also} amplified as it propagates back to the waveguide, and leaves the waveguide with a net amplification factor  $e^{2A(N-1)}$.  The result is a SNR and quantum Fisher information which grow exponentially with system size even when the total intracavity photon number is held fixed.}
		\label{fig:Protocol}
	\end{figure*}
	
	\section{Exponential SNR and QFI enhancement}
	\label{sec:ExpEnhancement}
	We now focus on computing the optimal SNR of the measurement operator $\hM_\tau(N)$ for our $N$ site chain in the $\epsilon \rightarrow 0$ limit; via Eq.~(\ref{eq:QFI}), this directly yields the QFI.  In this limit, a SNR $\sim 1$ will only be achieved for $\tau$ much longer than any internal dynamical timescale.  We thus consider the long-$\tau$ limit, effectively ignoring any transient behaviour and assuming the system is in its steady state.  
	Note that our system is dynamically stable as long as $w > \Delta$ and $\kappa>0$, ensuring that a steady state exists.
	
	From Eqs.~(\ref{eq:SignalDef}),(\ref{eq:Meas_Op}) and (\ref{eq:In_Out}), the first order in $\epsilon$ in this limit reads
	\begin{align}\label{eq:Signal_First_Order}
	    \mathcal{S}_\tau(N,\epsilon)
	    &
	    = 
	    \sqrt{2\kappa \tau} 
	    \big\lvert
	    \Re
	    [
	    e^{-i \phi}
        \delta 	\langle \ha_1 \rangle^{\rm ss}
	    ]
	    \big\rvert
	\end{align}
	where
	\begin{equation}
	    \delta \langle \ha_1 \rangle^{\rm ss} \equiv    
	        \epsilon \lim_{\epsilon \rightarrow 0}
	        \left( 
	        \frac{\langle \ha_1 \rangle^{\rm ss}_\epsilon - \langle \ha_1 \rangle^{\rm ss}_0}
	        {\epsilon} \right)
	\end{equation}
	is the steady state linear response of the site-$1$ average amplitude to a non-zero $\epsilon$.  This response will be determined by the zero-frequency susceptibilities (Green's functions) of the unperturbed system.  
	
    It will be convenient to split up $\delta \langle \ha_1 \rangle^{\rm ss}_\epsilon$ into its real and imaginary parts, or  equivalently to think of the dynamics in the quadrature picture. There are then four different types of  susceptibilities:  $\chi^{\alpha \beta}[n,m;\omega]$ is the response of the $\alpha$ quadrature on site $n$ to a force which directly drives the $\beta$ quadrature on site $m$. From Eqs.~(\ref{eq:X_EOM})-(\ref{eq:P_EOM}) and Eq.(\ref{eq:Susceptibility}), we find that the $\epsilon=0$ susceptibilities are 
	\begin{align}
	    \chi^{xx}[n,m;\omega] \label{eq:chi_x}
	    &=
	    e^{A(n-m)} \tilde{\chi}^{xx}[n,m;\omega],
	    \\ \label{eq:chi_p}
	    \chi^{pp}[n,m;\omega]
	    &=
	    e^{-A(n-m)} \tilde{\chi}^{pp}[n,m;\omega],
	    \\ \label{eq:Off_Diag}
	    \chi^{xp}[n,m;\omega]
	    &=
	   \chi^{px}[n,m;\omega]
	   =
	   0.
	\end{align}
	Here $\tilde{\chi}^{\alpha \beta}[n,m;\omega]$ the susceptibility of a Hermitian $N$ site tight-binding chain with hopping $i J$ and amplitude decay rate $\kappa/2$ on the first site (see Appendix \ref{app:chi_particle}). The above structure reflects the fact that the dynamics of the $\hx$ and $\hp$ quadratures correspond to two uncoupled copies of the Hatano-Nelson chain with opposite signed imaginary vector potential $A$. Hence, $\hx$ quadrature signals are amplified as they propagate to the right, and deamplified as they traverse to the left, while the opposite is true for $\hp$ quadrature signals.  Note that if we started with two explicit Hatano-Nelson chains, the discussion here would be identical; $x$ and $p$ would then just index the two different chains.   
	
	To proceed, we need to specify the form of the perturbation Hamiltonian $\hat{V}$.  Our system exhibits the the non-Hermitian skin effect (NHSE), implying a strong sensitivity to changes in boundary conditions.  As the unperturbed system is an open chain, this suggests that an optimal $\hat{V}$ would induce tunneling between the first and last site, i.e. 
	\begin{align}
	    \hat{V}_{\rm NHSE} 
	    = 
	   e^{i \varphi}\ha^\dagger_1 \ha_N 
       +    
       e^{-i \varphi }\ha^\dagger_N \ha_1,
	\end{align}
	with $\varphi$ an arbitrary phase. As we show in Appendix \ref{app:SNR}, this choice of $\hat{V}$ does not result in an enhanced sensitivity if one uses the proper metric of QFI/$\bar{n}_{\rm tot}$ (or equivalently SNR/$\sqrt{\bar{n}_{\rm tot}}$).  While the signal produced by $\hat{V}_{\rm NHSE}$ is large, this is simply because our system is an amplifier with a large end-to-end gain. The number of photons on the last site (and hence $\bar{n}_{\rm tot}$) will be amplified equally by this gain.  As a result, QFI/$\bar{n}_{\rm tot}$ does not show any enhancement as one increases the system size $N$, nor any enhancement over a conventional, single-cavity dispersive detector.  We are thus left with a depressing conclusion: the non-Hermitian skin effect does not provide any true advantage in sensing. Note also that $\hat{V}_{\rm NHSE}$ does not break the $\mathbb{Z}_2$ symmetry of the unperturbed system (see App.~\ref{app:Symmetry}).

    Luckily, this is not the end of the story.  Enhanced sensing is possible with our system, if we chose a $\hat{V}$ that fully exploits the opposite chiralities of our two (effective) Hatano-Nelson chains. Consider the innocuous-looking purely local perturbation
	\begin{align}
	    \hat{V}_{N} = \ha_N^\dagger \ha_N,
	\end{align}
	so that $\epsilon$ now corresponds to a small change in the resonance frequency of the last site. 
	This perturbation does indeed break the $\mathbb{Z}_2$ symmetry of the unperturbed system.   To understand how $\hat{V}_{N}$ affects the dynamics of the lattice, it is best to re-examine the equations of motion in the $\hx$ and $\hp$ basis. They remain the same everywhere except the last site $N$, where they now read
	\begin{align}
	\label{eq:XPerturbation}
	&\dot{\hx}_N 
	=
	J e^{A} \hx_{N-1} + \epsilon \hp_{N},
	\\
	\label{eq:PPerturbation}
	&\dot{\hp}_N
	=
	Je^{-A} \hp_{N-1}-\epsilon \hx_N.
	\end{align}
	Recall that without the perturbation present, the dynamics of the $\hx$ and $\hp$ quadratures are completely independent (c.f.~Eqs.~(\ref{eq:X_EOM}) and (\ref{eq:P_EOM})). The dispersive shift $\epsilon$ on site $N$ now effectively couples the two non-Hermitian chains, thereby breaking phase-dependent non-reciprocity (see Fig.~\ref{fig:Schematic}). While the intuitive picture of directional amplification remains unchanged in the rest of the lattice, a wavepacket with a well defined global phase can now scatter off of the perturbation $\epsilon$ and change its phase in the process. The role of $\epsilon$ is reminiscent to that of a magnetic impurity in the quantum spin Hall effect: in both cases the propagation direction of a particle is determined by some internal degree of freedom, which the impurity can change \cite{Spin_Hall_Review}.

	We next judiciously choose the phase of the drive $\beta$ to be real and the measurement angle $\phi = \pi/2$. Equivalently, we apply a driving force $-\sqrt{2\kappa} |\beta|$ to $\hx_1$ and consider the corresponding response of its canonically conjugate quadrature $\hp_1$. When $\epsilon = 0$, this off-diagonal susceptibility vanishes, see Eq.(\ref{eq:Off_Diag}).  To first order in $\epsilon$, it becomes non-zero.
		We further take $N$ to be odd in what follows, as this guarantees (via the chiral symmetry of our unperturbed system) that the lattice will have a resonant mode at zero frequency.  This then provides a further resonant enhancement of our system's zero frequency response properties.  Note that for an even $N$, we would still have the same exponential enhancement quoted in Eqs.~(\ref{eq:chi_x})-(\ref{eq:chi_p}); in this case however, there is no resonant mode at zero frequency, causing a suppression of susceptibilities by a multiplicative factor of $\kappa/(2J)$ (see Eqs.(\ref{eq:chi_n1})-(\ref{eq:chi_1n})).  
		
	With these optimized choices, first order perturbation theory yields: 
	\begin{align}
	    \nonumber
	    \mathcal{S}_\tau(N,\epsilon)
	    &
	    =
	    \sqrt{\kappa \tau}
	    |\sqrt{2 \kappa} \beta|
	    \Big(
	    |\delta \chi^{px}[1,1;0]|
	    \Big)
	    \\ \nonumber 
	    &
	    =
	    \sqrt{2\kappa \tau}\sqrt{\kappa}|\beta|
	    \Big(
	    |\chi^{pp}[1,N;0]\epsilon \chi^{xx}[N,1;0]|
	    \Big)
	    \\ \label{eq:Large_SNR0}
	    &
	    =
	    \sqrt{8\kappa \tau \bar{n}_{N}}
	    \left|
	    \frac{\epsilon}{\kappa}
	    \right|
	    e^{A(N-1)}.	    
	\end{align}
	Here $\bar{n}_N$ denotes the leading-order-in-$\beta$ average photon number of the last site in the lattice, and is given by:
	\begin{equation}
	    	\bar{n}_N = |\langle \ha_N \rangle^{ss}_0|^2 = \kappa|\beta|^2 |\chi^{xx}[N,1,0]|^2 \propto e^{2A(N-1)}
	\end{equation}
    For large $A$, the average photon number on site $N$ is exponentially larger than that on other sites.  Writing
    $\bar{n}_N = Z(A) \bar{n}_{\rm tot}$
    we have
    $ Z(A) =  1 -  \mathcal{O}(e^{-4A})$ (see App. \ref{app:n_bar}). We thus obtain:
    \begin{equation}
        \mathcal{S}_\tau(N,\epsilon)
	    = 	    
	    \sqrt{8 Z(A) \kappa \tau  \bar{n}_{\rm tot}}
	    \left|
	    \frac{\epsilon}{\kappa}
	    \right|
	    e^{A(N-1)}.	    
	    	\label{eq:Large_SNR}
    \end{equation}
    Eq.~(\ref{eq:Large_SNR}) is a central result of this work:  it shows that even when the total photon number
    $\bar{n}_{\rm tot}$ is held fixed, our system exhibits a signal power that grows exponentially with system size.  
    
	For an intuitive picture, consider the propagation of $x$-quadrature photons injected from the waveguide into site $1$, as depicted in Fig.~\ref{fig:Protocol}.  These photons will propagate to the last site $N$, with an amplitude $\chi^{xx}[N,1;\omega] \propto e^{A(N-1)}$. Photons that then scatter off the perturbation $\epsilon \hat{V}_N$ will change phase, so that they now correspond to the $p$ quadrature (c.f.~Eq.~(\ref{eq:PPerturbation})). They can then propagate back to the first lattice site with an amplitude $-\epsilon\chi^{pp}[1,N;\omega] \propto e^{A(N-1)}$.  This simple scattering process (involving both $x$ and $p$ quadrature propagation) leads to a parametrically large signal in  $\hp_1$.  
	
	The above heuristic picture also explains why the signal is amplified more than the average photon number $\bar{n}_{\rm tot}$:  the average photon number only involves amplification along one traversal of the chain, whereas the signal magnitude involves two traversals (forward and back).  This directly explains the extra large factor of $e^{A(N-1)}$ in Eq.~(\ref{eq:Large_SNR}).  We stress that this exponential signal enhancement would also occur in dissipative realizations of our doubled Hatano-Nelson chain.
	
	The final step in characterizing our sensor is to examine its noise properties.  Naively, one might expect that the same dynamics responsible for our signal enhancement would also exponentially amplify fluctuations in the output field.  This is not the case: as already discussed, calculating the QFI only requires computing the noise to zeroth order in $\epsilon$, see Eq.(\ref{eq:QFI}). Without the perturbation, the two effective Hatano-Nelson chains are completely decoupled.  Thus, 
	any noise entering through the waveguide will undergo equal amounts of amplification and deamplification before exiting the lattice.  For the ideal case of zero internal loss, this means that the noise temperature of the output field will be identical to that of the input field. As a result, the noise in the homodyne current is
    \begin{align}\label{eq:Noise}
	\mathcal{N}_{\tau}(N, 0) = \sqrt{\bar{n}_{\rm th}+\frac{1}{2}}
	\end{align}
	with $\bar{n}_{\rm th}$ representing the number of thermal quanta in the input field.  	

	Combining these two results, our signal-to-noise ratio is 
	\begin{align} \nonumber
	\text{SNR}_\tau(N,\epsilon)
	&=
	4
	\sqrt{\frac{Z(A) \bar{n}_{\rm tot}\kappa \tau}{2\bar{n}_{\rm th}+1}} |\frac{\epsilon}{\kappa}| e^{A(N-1)}
	\\  \label{eq:LargeSNR}
	&=
	\sqrt{Z(A)}
	e^{A(N-1)} \,
	\text{SNR}_\tau(1,\epsilon),
	\end{align}
	where $\text{SNR}_{\tau}(1,\epsilon)$ is the signal-to-noise ratio of a ubiquitous single-mode dispersive detector \cite{RMP_Clerk,Ben_PRX}. As we have stressed, $\text{SNR}_\tau(N,\epsilon)$ also represents the QFI of our system.
	We see that the SNR and QFI can be exponentially enhanced by either increasing system size $N$ or amplification factor $A$, while all the while maintaining a {\it fixed} total photon number $\bar{n}_{\rm tot}$.  This is the central result of our work. The crucial ingredients here are the inherent chiral amplification present in a Hatano-Nelson chain, the effective symmetry breaking that occurs when coupling the two opposite-chirality chains in our sensor, and the lack of any amplified output noise in the unperturbed system.
	
	Several comments are in order.  First, note that the large SNR achieved here is not contingent on approaching a parametric instability: our system is dynamically stable for any value of $\epsilon$ and $A$ (see Appendix \ref{app:Non-Pert}).  Second, the mechanism we discuss here is useful even in small systems, as the fixed photon number QFI has an exponential dependence on $A$; an arbitrarily large QFI can thus be achieved with only three lattice sites.  We further emphasize that the spatially-dependent amplification is a crucial aspect of our scheme. Indeed, the signal-to-noise ratio for a single-mode cavity amplifier can never achieve this sort of sensing enhancement, since the signal and noise are amplified in a similar manner \cite{Ben_PRX}. Finally,  we stress that this enhanced QFI in no way requires or is even related to the existence of an exceptional point in our dynamical matrix.    

	It is also worth stressing that our mechanism is completely distinct from other recently introduced methods that use parametric amplifiers to enhance dispersive sensing  \cite{Terhal_2014, Ben_PRX, Florian_PRL_2015}. These works exploit noise squeezing as the basic mechanism for enhancing the SNR and QFI.  Unfortunately, in many practical settings this squeezing is difficult to exploit, as one becomes extremely sensitive to the added noise of amplification stages that follow the primary measurement (i.e.~one needs following amplifiers to be quantum limited).  In contrast, our scheme does not rely on squeezing the measurement noise, but instead effectively amplifies the signal power at fixed total photon number.  The output noise has the same magnitude as the input noise, and hence taking advantage of our enhanced QFI does not need following amplification stages to be quantum limited.  This represents a significant practical advantage.    
	
    We end this section by pointing out that the $N$ dependence of the QFI in Eq.~(\ref{eq:LargeSNR}) does not violate standard Heisenberg-limit constraints \cite{Giovannetti2011}, as the setting here is different.  The usual Heisenberg limit applies to $N$ sensor systems which each interact independently with the parameter of interest; the QFI here scales as best as $\propto N$, a result which requires entanglement.    In contrast, each of the $N$ modes in our system is not an independent sensor interacting independently with the dispersive perturbation, as the sites are coupled.  The enhanced scaling we find is not the result of entanglement:  we stress that the input light to our system is just a coherent state.  Instead, the enhancement is a consequence of our system's unusual mechanism for non-reciprocal amplification. 
	
	\section{Non-Markovian Effects}
	
	We now relax the assumption that the parameter $\epsilon$ is infinitely weak.  For concreteness, we assume the sensing target is to distinguish the case $\epsilon = 0$ from 
	the case $\epsilon = \epsilon_0 \neq 0$.  This kind of discrimination is relevant in many practical situations, for example the dispersive measurement of the state of a qubit \cite{RMP_Clerk}.  We assume that $\epsilon_0$ is small enough such that linear response is still valid, but not so small that measurement will be infinitely long compared to internal system timescales.  We thus need to understand the finite-frequency response and noise properties of our non-Hermitian lattice sensor. 
	
    In this section, we will characterize our sensor by its measurement time $\tau_M$:  what is the minimum integration time to to achieve a SNR of unity?  Heuristically, $\tau_M$ is the minimum amount of time required to distinguish between $\epsilon = 0$ and $\epsilon = \epsilon_0$. In the limit $\epsilon_0 \rightarrow 0$, $\tau_M$ will be much longer than any internal sensor timescale, and we can use the long-time limit SNR expression derived in the previous section (c.f.~Eq.~(\ref{eq:LargeSNR})).  We define 
    $\tau^{*}_{M}(N)$ to be this $\epsilon_0 \rightarrow 0$ expression for the measurement time. Assuming that the input field has only vacuum noise, we find:
	\begin{align}\label{eq:T_meas_infty}
    	\tau^{*}_{M}(N)
	        =
	    \frac{1}{16 Z(A) \Bar{n}_{\rm tot} \kappa}
	    \left(\frac{\kappa}{\epsilon_0} \right)^2
	    e^{-2A(N-1)}.
	\end{align}
	The obviously attractive feature here is the exponential reduction of $\tau_M$ with increasing lattice size $N$ (but at fixed total photon number).
	
	As $N$ or $\epsilon_0$ is increased, $\tau^*_M(N)$ will become increasing smaller, and at some point will become comparable to internal system timescales.  At this point, the long-time limit assumption used to derive this expression becomes invalid.  There are two distinct relevant timescales that govern the  dynamics of our sensor.  The first $t_{rt}(N)$ determines the ballistic propagation time to traverse the lattice end to end:
	\begin{align}
    	 t_{rt}(N) = \frac{N}{J},
	\end{align}
	The second $t_{esc}(N)$ involves the coupling to the waveguide:  how quickly
	does a particle that is delocalized in the lattice leak out to the waveguide.  A simple Fermi's Golden Rule estimate yields the scale:
	\begin{align}
	    t_{esc}(N) = \frac{N+1}{\kappa}
	\end{align}
	Both these timescales increase with system size.  
	As a result, non-Markovian effects associated with internal dynamics become increasingly important with increasing $N$.  The crucial question is how this physics modified or places a limit on the exponential-in-$N$ measurement enhancement predicted by Eqs.~(\ref{eq:LargeSNR}) and 
	(\ref{eq:T_meas_infty}).
	For large enough $N$ the measurement will be so fast that these internal timescales matter.  
	Do they simply put a bound on the measurement time, or does performance continue to increase with increasing $N$?
	
	We first consider the limit $J \gg \kappa$; the only relevant dynamical timescale is then $t_{esc}(N)$, the time it takes a delocalized photon to escape the lattice. In this regime, the level spacing of lattice resonances is much larger than their widths.  We can thus accurately approximate the relevant low-frequency behaviour of lattice susceptibilities by the contribution from the zero-frequency resonance (whose width is $1/t_{esc}(N)$).
	Assuming as always that $N$ is odd, we have:
    \begin{align}
    &
    \chi^{xx}[N,1;\omega]
    \approx
    \frac{2i^{N}}{N+1}
    \frac{e^{A(N-1)}}
    {\omega+i \frac{\kappa}{N+1}}
    \\
    &
    \chi^{pp}[1,N;\omega]
    \approx
    \frac{-2i^{-N}}{N+1}
    \frac{e^{A(N-1)}}
    {\omega+i \frac{\kappa}{N+1}}.
\end{align}
Note crucially that the residue at the poles are exponentially large in system size; this directly reflects the amplification physics we have discussed previously.  Because of these factors, the above response functions are not simply equivalent to those of a single mode system with a very small linewidth.  
    
With this approximation, we find that the SNR is given by (see Appendix \ref{app:Single_Pole} for details)

\begin{align} \label{eq:SNR_Approx}
	    &\text{SNR}_\tau(N, \epsilon,J \rightarrow \infty)
	        =
	    \\ 
	    &
	        \sqrt{ \frac{\tau}{\tau^*_{M}(N) }}
	        \left(
	            1+e^{-\frac{\tau}{t_{esc}(N)}}
	            -\frac{2 t_{esc}(N)}{ \tau }
	            (1-e^{-\frac{\tau}{t_{esc}(N)}})
	    \right)
	\nonumber
\end{align}
    The bracketed factor represents the non-Markovian correction to the long-time limit expression.  Note that the correction is only to the magnitude of the signal.  As we continue to use linear response, we only need to compute the homodyne current noise to zeroth order in $\epsilon_0$.  This noise is thus always vacuum noise regardless of the choice of integration time $\tau$.  
    
	 Using the above expression, we can then directly compute the 
	 measurement time $\tau_M$ in the $J \to \infty$ limit.  While finding the measurement time analytically is unfeasible, we can describe its asymptotic behavior in the strong and weak measurement limit (see App. \ref{app:Single_Pole})
	 \begin{align} \nonumber
		&\tau_M^{J = \infty}(N)
		=
		\begin{cases}
		\tau_M^*(N),
		& 
		 \tau_M^*(N) \gg t_{esc}(N)
		\\
       \sqrt{6}t_{esc}(N)
       \sqrt[\leftroot{-2}\uproot{2}5]{\frac{\tau_M^*(N)}{\sqrt{6}t_{esc}(N)}},
		& 
		 \tau_M^*(N) \ll t_{esc}(N)
		\end{cases}
		\\
		&
		\propto
		\begin{cases}
		e^{-2A(N-1)},
		&
		 \tau_M^*(N) \gg t_{esc}(N)
		\\
		(N+1)^{4/5}e^{-2A(N-1)/5},
		&
		\tau_M^*(N) \ll t_{esc}(N).
		\end{cases}
	 \end{align}
    We thus find a surprising result:  even for fast measurements where the escape time from the lattice plays a role, the measurement time continues to improve exponentially with lattice size $N$. Intuitively, this is because the deleterious effects of increasing the escape time $t_{esc} = (N+1)/\kappa$ with increasing $N$ is more than offset by the exponentially large number of photons $e^{2A(N-1)}$ that exit through the waveguide when $\epsilon = \epsilon_0$.
	
    We next consider the case where the hopping amplitude $J$ is not infinitely
    larger than all other scales.  In this case, we must also take into account the finite propagation speed  $v \propto J$ of particles the lattice. Because an injected wavepacket must make a round trip before acquiring any information about the perturbation $\epsilon$, for times less than $2N/v = N/J = t_{rt}(N)$ we expect the signal to be approximately zero. After this first round trip, the limiting factor in obtaining a large signal is once again the escape rate. 
    Including the effects of a finite $J$, we find that the SNR is well approximated by simply adding a cutoff to the $J \rightarrow \infty$ result in Eq.~(\ref{eq:SNR_Approx}):
    \begin{align} 
        \label{eq:SNR_Approx_J_Finite}
    	\text{SNR}_\tau(N,\epsilon, J)
    	& = 
            \Theta \left(\tau - t_{rt}\right )	
            \text{SNR}_\tau(N,\epsilon, J\rightarrow \infty)
 \end{align}
   where $\Theta(t)$ is the Heaviside step function.  This form reflects the basic intuition that it is impossible to make a measurement faster than the propagation time.

   Combining these results, we finally find that including the effects of both internal timescales $t_{rt}(N)$ and $t_{esc}(N)$, the measurement time (to good approximation) is given by
   \begin{align}\label{eq:meas_time}
       \tau_M(N)
       =
      \max
      (
      \tau_M^{J = \infty}(N), t_{rt}(N)
      )
   \end{align}
   where $\tau_M^{J = \infty}(N)$ is given in Eq.~(\ref{eq:T_meas_infty}).  Thus, as a 
   function of increasing system size $N$, the measurement time first decreases exponentially until it reaches the round-trip time in the lattice, after which it increases with $N$. The upshot of our analysis is that increasing the lattice size still provides an exponential sensing advantage when including non-Markovian effects.  This continues to be true until the measurement time is reduced to being on par with the  round-trip propagation time $t_{rt}(N) = N/J$. 
  
\begin{figure}[t]
	\centering
	\includegraphics[width=0.45\textwidth]{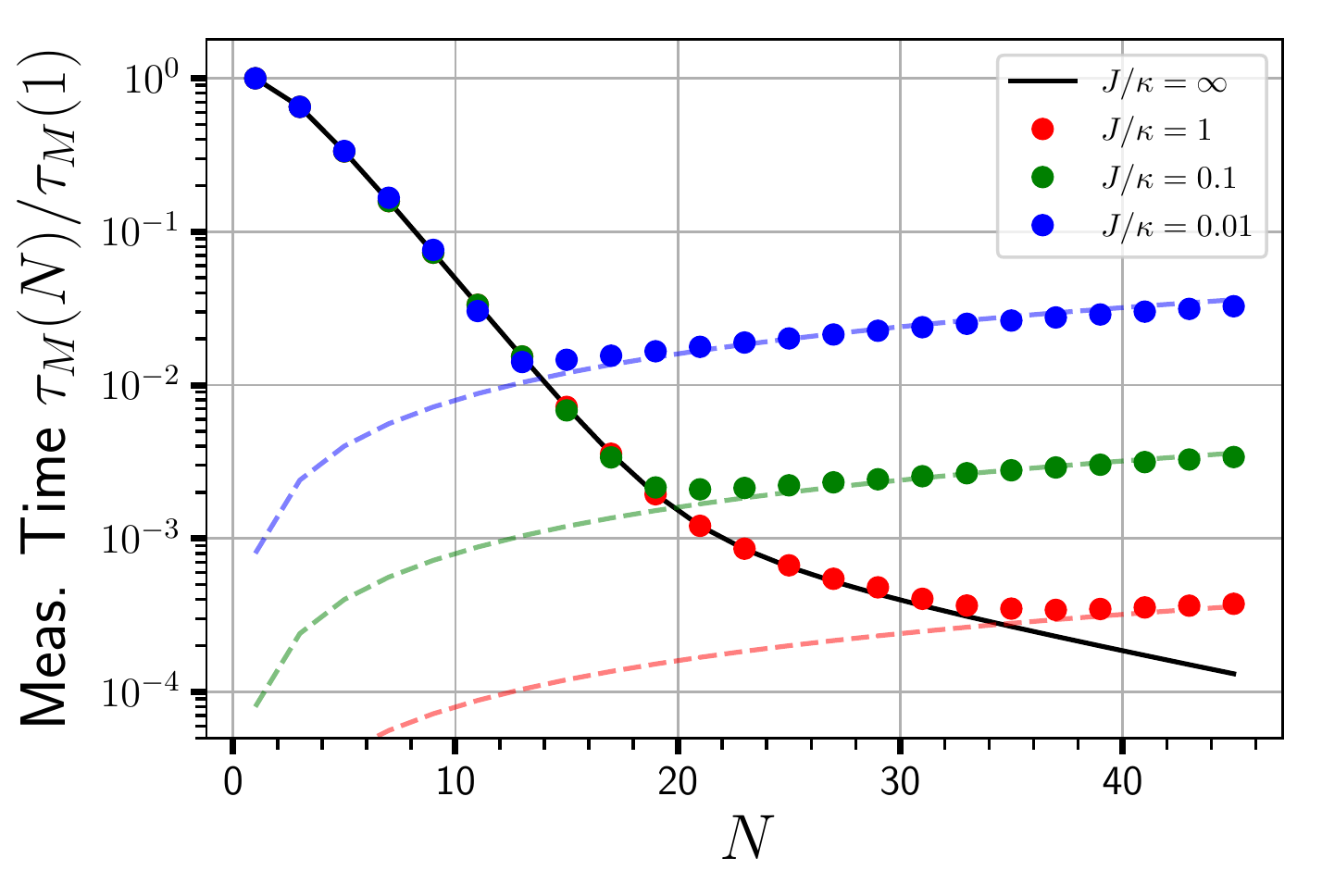}
	\caption{Measurement time $\tau_{M}(N)$ versus lattice size $N$, for different choices of the hopping amplitudes $J$.  The solid black line is the measurement time in the $J \rightarrow \infty$ limit,  $\tau^{J=\infty}_M(N)$. Faint dashed lines are the round trip 
	propagation timescale $t_{rt}(N) \equiv N/J$. The measurement time decays exponentially
	with increasing $N$, up until $\tau^{J = \infty}_{M}(N) \approx t_{rt} (N)$.  Further increases of $N$ cause the measurement time to scale with $t_{rt}(N)$, implying that it increases linearly with $N$. We take $\epsilon_0 = 10^{-8} \kappa$, $\Bar{n}_{\rm tot} = 5\times 10^{9}$ and $A = 0.2$.
	We also plot results for odd values of $N$ only, as this guarantees the existence of a zero-frequency lattice eigenstate and thus an additional resonant enhancement of our measurement (c.f.~main text before Eq.~(\ref{eq:Large_SNR0})).
	}
	\label{fig:Measurement_Time}
	\end{figure}

   In Fig.~\ref{fig:Measurement_Time}, we plot the numerically-calculated measurement time $\tau_M(N)$ versus lattice size $N$ for a fixed total photon number $\bar{n}_{\rm tot}$ and perturbation size $\epsilon_0 / \kappa$; different curves correspond to different values of the hopping $J / \kappa$.  We find an excellent agreement with the analytic approximation given in Eq.~(\ref{eq:meas_time}).
    The measurement time follows $\tau_M^{J = \infty}(N)$ (dark solid line) until it reaches the round-trip time $t_{rt}(N)$ (faint dashed lines), after which it increases linearly with $N$.
   
    \section{Beyond linear response}\label{sec:Beyond_Linear_Response}
	In this final section, we again consider the sensing problem of distinguishing $\epsilon = 0$ from $\epsilon = \epsilon_0$; now however, we  analyze the regime where (due to amplification effects) $\epsilon_0$ is too large for a linear response analysis to be valid. This is in contrast to the previous section, where $\epsilon_0$ was small enough that linear response was still valid, but large enough that non-Markovian detector effects were important.  
	
    For any $\epsilon_0$ the output state of the light leaving the waveguide will be Gaussian, and the statistics of the measured homodyne current will be Gaussian.  We can thus again quantify our sensor's performance by calculating the signal-to-noise ratio.  We now however need to account for the fact that the homodyne current noise will also depend on $\epsilon_0$. The definition of the signal-to-noise ratio becomes:
	\begin{align}\label{eq:SNR_Average}
	    \mathrm{SNR}_\tau(N,\epsilon_0)
	    \equiv
	    \frac
	    {
	    |
	    \langle \hM_\tau(N) \rangle_{\epsilon_0}
	    -
	    \langle \hM_\tau(N) \rangle_{0}
	    |
	    }
	    {\sqrt{
	    \frac{\mathcal{N}^2_{\tau}(N,0)+\mathcal{N}^2_{\tau}(N, \epsilon_0)}
	    {2}}}
	\end{align}
    This SNR quantifies the distinguishability between the Gaussian homodyne current distributions obtained for $\epsilon = 0$ versus $\epsilon = \epsilon_0$ 
    (see e.g.~\cite{RMP_Clerk, Laflamme_PRL}). 
    
    As might be expected, the nonlinear dependence of SNR on $\epsilon_0$ will 
    prevent one from indefinitely improving the measurement with increasing $N$.  The key issue is that beyond linear response, noise amplification will also play a role.   We show in what follows that even with this complication, our system yields a strong advantage, allowing one to fundamentally change the scaling of the SNR with $\epsilon_0$. 
    
    We will focus on the most interesting situation where $\epsilon_0/\kappa \ll 1$, but where linear response breaks down because of a large amplification factor (i.e.~$e^{A(N-1)} \epsilon_0/\kappa $ is not necessarily small). Further, we take the round-trip time $t_{rt}(N) = N/J$ to be small enough that we can ignore the transient dynamics and consider only the steady-state response.  Formally, we now need 
    to calculate the output field leaving the waveguide to all orders in $\epsilon_0$. We thus expand the zero frequency quadratures of the output field as a power series in $\epsilon_0 / \kappa$:
    \begin{align}
        \hat{X}^{(\rm out)}[0] & \equiv 
            \sum_{k=0}^{\infty} 
            \left( \frac{\epsilon_0}{ \kappa} \right)^k
            \hat{X}^{(\rm out)}_k \\
        \hat{P}^{(\rm out)}[0] & \equiv 
            \sum_{k=0}^{\infty} 
            \left( \frac{\epsilon_0}{ \kappa} \right)^k
            \hat{P}^{(\rm out)}_k 
    \end{align}
    To zeroth order in $\epsilon_0$, there is no mixing of quadratures, and input signals are reflected with no net amplification (but just a trivial sign change):
    \begin{align}
        \hat{X}^{(\rm out)}_0 = - 
            \hat{X}^{(\rm in)}[0], 
        \,\,\,
        \hat{P}^{(\rm out)}_0 = 
            - \hat{P}^{(\rm in)}[0], 
    \end{align}
    Note that throughout this section, we associate the coherent drive tone amplitude $\beta$ with the average value of $ \hat{X}^{(\rm in)}[0]$.
    
    In contrast, the first order contributions correspond to a process where input fields scatter once off the ``impurity" before returning to the waveguide.  This scattering converts one canonical quadrature to the other, and also results in a net amplification or deamplification
   \begin{align}
        \hat{X}^{(\rm out)}_1 
            & = 
            4e^{-2A(N-1)} \hP^{\rm(in)}[0] \\
        \hat{P}^{(\rm out)}_1 
            & = 
             -4 e^{2A(N-1)}
	    \hX^{\rm (in)}[0]
    \end{align}
    The amplification of $\hat{X}^{(\rm in)}$ is exactly the process we discussed in 
    Sec.~\ref{sec:ExpEnhancement} that is responsible for the exponentially-enhanced signal.  The attenuation of $\hat{P}^{(\rm in)}$ at this order can be understood analogously.  
    
    What about the second order in $\epsilon_0$ contributions?  Heuristically, these correspond to input fields scattering off the impurity twice.  While we expect such a process to preserve the identity of each canonical quadrature, it also has a more surprising feature: it results in no net amplification or deamplification:
  \begin{align}
        \hat{X}^{(\rm out)}_2 
             = 
            8 
            \hX^{\rm(in)}[0], 
            \,\,\,\,\,\,
        \hat{P}^{(\rm out)}_2 
             = 
              8 
	        \hP^{\rm (in)}[0]
    \end{align}
    This unexpected result can again be traced by to the chiral and quadrature-dependent nature of gain and loss in our system.  Interacting with 
    the impurity twice implies that an input signal has performed at least two round-trip traversals of the lattice (partially as an $X$, partially as a $P$).  The gain and attenuation for each of these roundtrips necessarily cancel. 
    
    This pattern continues to higher order, and provides a simple explanation for the full expression we find for the output field:  the net amplification / deamplification factor for each kind of quadrature to quadrature scattering process is independent of $\epsilon_0$.  We find
    \begin{align}
	   &\hX^{\rm (out)}[0]
	   =
	   R(\epsilon_0)\hX^{\rm (in)}[0]
	   -T(\epsilon_0) e^{-2A(N-1)} \hP^{\rm (in)}[0]
	   \\
	   &\hP^{\rm (out)}[0]
	   =
	   T(\epsilon_0) e^{2 A(N-1)}
	   \hX^{\rm (in)}[0]
	   +
	   R(\epsilon_0) \hP^{\rm (in)}[0]	   
	\end{align}
	where
	\begin{align}
	    &
	    R(\epsilon_0)
	    =
	    -
	    \frac
	    {(\frac{\kappa}{2})^2-\epsilon_0^2}
	    {(\frac{\kappa}{2})^2+\epsilon_0^2}
	    \\
	    &
	    T(\epsilon_0)
	    =
	    \frac{\kappa \epsilon_0}
	    {(\frac{\kappa}{2})^2+\epsilon_0^2}
	\end{align}
	are elements of an orthogonal scattering matrix describing the conversion of quadratures (see Appendix \ref{app:Non-Pert} for details). 
	We see that quadrature-preserving scattering processes never come with amplification factors, whereas the amplification factors for quadrature-changing scattering are independent of $\epsilon_0$.  Crucially, there are no amplification factors in denominators in this expression.  This result can be derived via a canonical squeezing transformation which eliminates the anomalous terms in Eq.~(\ref{eq:HBKC}); it also reflects the fact that our system is dynamically stable regardless of the strength of $\epsilon_0$.
	
	From these input-output relations, we can readily compute the SNR.  Taking the noise of the input field to be vacuum, we have:
	\begin{align} \nonumber
	    \mathrm{SNR}_\tau(N,\epsilon_0)
	    =
	    &
	    \frac{
	    \sqrt{8 \tau}|\beta||T(\epsilon_0)|e^{2A(N-1)}
	    }
	    {
	    \sqrt{1+R^2(\epsilon_0)+T^2(\epsilon_0)e^{4A(N-1)}}
	    }
	    \\ \label{eq:Full_SNR}
	    =
	    &
	    \frac{
	    \sqrt{2 Q(A,\epsilon_0) \bar{n}_{\rm tot}\kappa \tau }|T(\epsilon_0)|e^{A(N-1)}
	    }
	    {
	    \sqrt{1+R^2(\epsilon_0)+T^2(\epsilon_0)e^{4A(N-1)}}
	    }
	\end{align}
	where $\bar{n}_{\rm tot} = (\bar{n}_{\rm tot}(0) + \bar{n}_{\rm tot}(\epsilon_0))/2$.

\begin{figure}[t]
	\centering
	\includegraphics[width=0.45\textwidth]{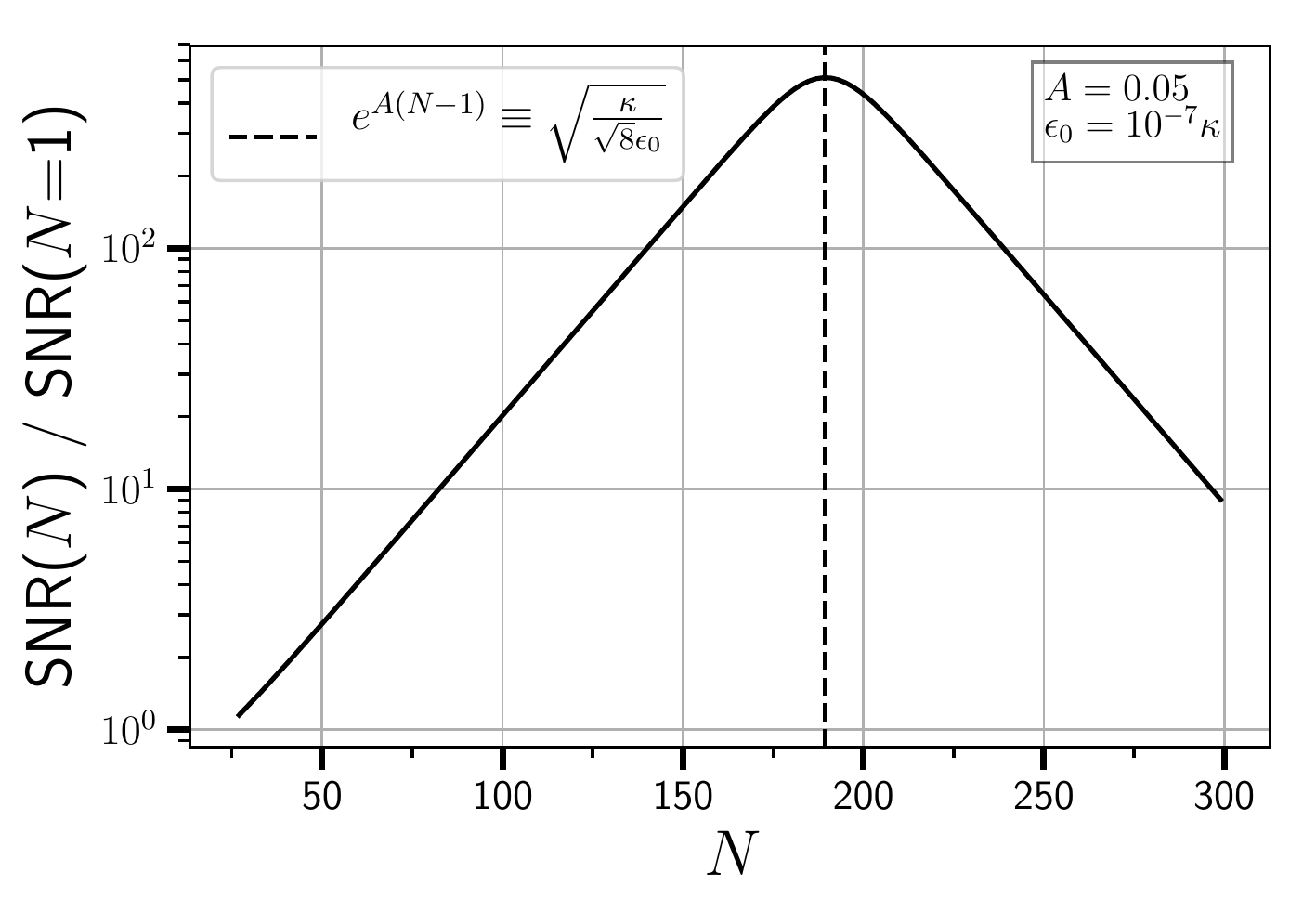}
	\caption{Non-perturbative signal-to-noise ratio in the long time limit $\text{SNR}_{\tau}(N, \epsilon_0)/\text{SNR}_{\tau}(1, \epsilon_0)$, as a function of lattice size $N$. The SNR initially increases exponentially with $N$, as predicted by our linear-response analysis in Sec.~\ref{sec:ExpEnhancement}. For sufficiently large $N$, linear response breaks down due to the amplification of noise; this causes the SNR to decrease with $N$ for large $N$.  A non-trivial maximum is thus reached for an intermediate value of $N$ given by Eq.~(\ref{eq:A_Opt}).  For this optimal $N$ and a weak perturbation $\epsilon_0$, the SNR scales like $\sqrt{\epsilon_0/\kappa}$ (as opposed to the more standard scaling $\epsilon_0/\kappa$) . The parameters here are $A = 0.05$, $\epsilon_0 = 10^{-7} \kappa$ and $\bar{n}_{\rm tot} = 5\times 10^{9}$. We only plot the results for odd values of $N$, which ensures an resonant enhancement of the zero-frequency response (c.f discussion preceding Eq.~(\ref{eq:Large_SNR0}))}
	\label{fig:Scatter_Twice}
	\end{figure}

	We see that now, the denominator in Eq.~(\ref{eq:Full_SNR}) also depends on the amplification factor $A$, which corresponds to the amplification of noise.  Because of this, increasing $A$ and/or $N$ indefinitely is no longer optimal.  There remains nonetheless an advantage in using a carefully chosen amount of amplification. Ignoring $Q(A, \epsilon_0)$ and maximizing the SNR Eq.~(\ref{eq:Full_SNR}) with respect to the amplification, we see that the optimal choice corresponds to amplification that simply doubles the output noise over pure vacuum noise.  In the $\epsilon_0 \ll \kappa$ limit of interest, the condition is:
    \begin{align}\label{eq:A_Opt}
	    e^{4A^*(N-1)}
	    \equiv
	    \frac{1+R^{2}(\epsilon_0)}{T^2(\epsilon_0)}
	    \approx
	    \frac{\kappa^2}{8 \epsilon_0^2}
	\end{align}
    With this optimized choice of $A$, the SNR  written in terms of $\epsilon_0$ is then
	\begin{align}
	    \mathrm{SNR}_\tau(N,\epsilon_0)
	    =
	    8^{1/4} \sqrt{Q(A^*,\epsilon_0)\bar{n}_{\rm tot} \kappa \tau}
	    \sqrt{\frac{\epsilon_0}{\kappa} }.
	\end{align}
	We show in Appendix \ref{app:Non-Pert} that $Q(A^*,\epsilon_0) =   1-\mathcal{O}((\frac{8 \epsilon_0^2}{\kappa^2})^{\frac{1}{N-1}})$. Comparing against Eq.~(\ref{eq:Large_SNR0}), we see that the optimized amplification has changed the fundamental scaling of the long-time SNR from being linear in the small parameter $\epsilon_0/\kappa$ to a square-root dependence.
	Thus, by extending our analysis beyond a simple linear-response treatment, we see that the exponential enhancement of the SNR predicted in Eq.~(\ref{eq:Large_SNR0}) cannot extend indefinitely:  the best one can do is to enhance the SNR (over a conventional dispersive measurement) by a large factor $\sqrt{\kappa / \epsilon_0}$.
	This behaviour is plotted in Fig.~(\ref{fig:Scatter_Twice}).
	We again note that this predicted measurement enhancement does not require a large number of lattice sites; just three is already enough.
	
    The enhanced square-root dependence of the SNR on $\epsilon_0$ is superficially reminiscent of the behaviour found in non-Hermitian exceptional point (EP) sensors \cite{Wiersig_2014}.  We stress that these phenomena are completely distinct.  For EP sensors, it is the frequency of a resonance that exhibits a square root dependence, and not the SNR of a specific measurement (or other metric that also quantifies fluctuations).  Further, EP sensing is based on operating near a point where the system's dynamical matrix becomes defective and normal modes coalesce.  In contrast, our system is not operating near such a special operating point.  As we have stressed, the mechanism for enhanced SNR in our system is based on its directional amplification, and its ability to amplify signals and noise differently.  
    
\section{Conclusion}

In this work, we have shown how the unique features of non-Hermitian lattice dynamics can be used for highly enhanced Hamiltonian parameter estimation and parametric sensing.  We analyzed a concrete setup involving two copies of the Hatano-Nelson model and a symmetry breaking perturbation.  The response to the perturbation grows exponentially with system size, even when the total system photon number is kept fixed. Our analysis focused on a specific realization of this idea using a chain of parametrically driven cavities and a standard dispersing coupling to the parameter of interest.  Here, even in the presence of quantum noise effects, the SNR and quantum Fisher information both grow exponentially with system size (all the while keeping photon number fixed).  The system we described could be achieved in a variety of superconducting circuit and quantum optical platforms, and only requires one to make a homodyne measurement of the output field leaving the sensor.  We also analyzed effects that go beyond standard linear-response and Markovian assumptions.  Even including higher-order effects, we show that our scheme allows one to dramatically enhance the SNR so that it depends on the square root of the sensing parameter.  

Our work highlights the usefulness of multi-mode non-Hermitian features that go beyond the mere existence of exceptional points.  An open question is whether other unique features attributed to non-Hermiticity, such as exotic topological phases or chiral mode switching, are also advantageous to quantum sensing problems. 

\section*{Acknowledgements}
We thank Kero Lau for a careful reading of the manuscript. This material is based upon work supported by the Air Force Office of Scientific Research under award number FA9550-19-1-0362.

\appendix
\section{$\mathbb{Z}_2$ symmetry of a non-Hermitian tight-binding model}\label{app:Symmetry}
We discuss in more detail the $\mathbb{Z}_2$ symmetry that we wish to break to order to obtain an exponentially large response. We consider two finite, $N$ site open Hatano-Nelson lattices with opposite chiralities
(i.e.~oppositely signed imaginary vector potentials $A$). The time-dependent Schr\"{o}dinger  equation reads
	\begin{align}
    	\dot{\psi}^{\uparrow}_n 
    	= 
    	J e^{A}  \psi^{\uparrow}_{n-1} 
    	- 
    	J e^{-A} \psi^{\uparrow}_{n+1}
    	\\
     	\dot{\psi}^{\downarrow}_n 
    	= 
    	J e^{-A}  \psi^{\downarrow}_{n-1} 
    	- 
    	J e^{A} \psi^{\downarrow}_{n+1}  
	\end{align}
where $\sigma$ indexes the two chains. These equations of motion are invariant under a combination of time reversal $\dot{\psi}^{\sigma}_n \to - \dot{\psi}^{\sigma}_n $, spatial inversion $\psi^{\sigma}_n \to \psi^{\sigma}_{N+1-n}$ and pseudospin inversion $\sigma \to \bar{\sigma}$.

While this is a seemingly trivial symmetry, we note that the Heisenberg equations of motion of our dissipation-free realization of the same model
	\begin{align}\label{eq:Heisenberg_x_App}
	\dot{\hx}_n & = 
	J e^{A} \hx_{n-1} - 
	J e^{-A} \hx_{n+1},
	\\ \label{eq:Heisenberg_p_App}
	\dot{\hp}_n & = 
	J e^{-A} \hp_{n-1} - 
	J e^{A} \hp_{n+1},
	\end{align}
are invariant under the same set of symmetries, where $\hx_n$ and $\hp_n$ play the role of pseudospin. To describe these symmetries requires using operators acting on the bosonic Hilbert space. To this end, we consider the antiunitary time reversal operator $\mathcal{T}$, a unitary rotation operator $\mathcal{R}$ and the unitary spatial inversion operator $\mathcal{S}$ whose action on the quadratures reads
\begin{align}
    \mathcal{T} 
    \hx_n
    \mathcal{T}^{-1}
    =
    \hx_n,
    &
    \hspace{0.25cm}
    \mathcal{T}
    \hp_n
    \mathcal{T}^{-1} 
    =
    -\hp_n
    \\
    \mathcal{R}
    \hx_n
    \mathcal{R}^{-1} 
    =
    \hp_n,
    &
    \hspace{0.25cm}
    \mathcal{R}
    \hp_n
    \mathcal{R}^{-1} 
    =
    -\hx_n
    \\
    \mathcal{S}
    \hx_n
    \mathcal{S}^{-1}
    =
    \hx_{N+1-n},
    &
    \hspace{0.25cm}
    \mathcal{S} 
    \hp_n
    \mathcal{S}^{-1}
    =
    \hp_{N+1-n}.
\end{align}
Equivalently, we have
\begin{align}
&\mathcal{T} \ha_n \mathcal{T}^{-1} = \ha_n
\\
&\mathcal{R} \ha_n\mathcal{R}^{-1} = -i\ha_n
\\
& \mathcal{S} \ha_n\mathcal{S}^{-1} = \ha_{N+1-n}
\end{align}
With these definitions in hand,  it is easy to verify that the Hermitian Hamiltonian which gives the equations of motion Eqs.(\ref{eq:Heisenberg_x_App}) and (\ref{eq:Heisenberg_p_App})
	\begin{align}
	   \hH_B
	   =
	   J
	\sum_{n=1}^{N-1}
    \left(
	-e^{-A}\hx_{n+1} \hp_n + e^{A} \hp_{n+1} \hx_n
	\right).
	\end{align}
is invariant under the combination of time-reversal, rotation, and spatial inversion. Similarly, the non-local perturbation  considered in Section \ref{sec:ExpEnhancement}
\begin{align}
    \hat{V}_{\rm NHSE}
    =
    e^{i \varphi} \ha_1^\dagger \ha_N
    +
    e^{-i\varphi} \ha_N^\dagger \ha_1
\end{align}
is invariant under the same combination of symmetries. Time-reversal changes the phase $\varphi \to -\varphi$, $\hat{V}_{\rm NHSE}$ commutes with $\mathcal{R}$ and spatial inversion sends $\ha^\dagger_1 \ha_N \to \ha_N^\dagger \ha_1$.

\section{Quadrature susceptibility matrices}\label{app:chi_quadrature}
We first compute the susceptibilities for the $\hx_n$ and $\hp_n$ quadratures, defined as $\ha_n = (\hx_n+i\hp_n )/\sqrt{2}$. The Heisenberg-Langevin equations of motion in this basis are
	\begin{align}\label{eq:HL_x}
	\dot{\hx}_n & 
	= 
    -i[\hx_n, \hat{H}_B]
    -
    \delta_{n1}
    \left(
    \frac{\kappa}{2}\hx_n
    +
    \sqrt{\kappa}
    \left(
    \sqrt{2}\beta + \hX^{\rm(in)}
    \right)
    \right)
    \\
    \label{eq:HL_p}
	\dot{\hp}_n & = 
    -i[\hp_n, \hat{H}_B]
    -
    \delta_{n1}
    \left(
    \frac{\kappa}{2}\hp_n
    +
    \sqrt{\kappa}
    \hP^{\rm(in)}
    \right),
	\end{align}
	where $\hX^{\rm (in)}$ and $\hP^{\rm (in)}$ are the operator equivalent of Gaussian white noise. They average to zero, and their second moment is
	\begin{align}
	    &
	    \langle 
	    \hX^{\rm (in)}(t) \hX^{\rm (in)} (t')
	    \rangle
	    =
	    \left(
	    \bar{n}_{\rm th}
	    +
	    \frac{1}{2}
	    \right)
	    \delta(t-t')
	    \\
	    &
	    \langle 
	    \hP^{\rm (in)}(t) \hP^{\rm (in)} (t')
	    \rangle
	    =
	    \left(
	    \bar{n}_{\rm th}
	    +
	    \frac{1}{2}
	    \right)
	    \delta(t-t')
	    \\
	    &
	    \frac{1}{2}
	    \langle
	    \{
	    \hX^{\rm (in)}(t), \hP^{\rm (in)}(t')
	    \}
	    \rangle
	    =
	    0
	\end{align}
	where $\bar{n}_{\rm th}$ is the number of thermal quanta in the input field. We now focus on the case where $\bar{n}_{\rm th} = 0$, with generalizations to finite-temperature inputs being straightforward. 
	
	An immense simplification arises by making a local Bogoliubov (squeezing) transformation, such that the Hamiltonian preserves the total number of these new quasiparticles. The dynamical matrix of in this new basis is then explicitly Hermitian. Defining new canonically conjugate quadrature operators $\htx_n$ and $\htp_n$ by
	\begin{align}\label{eq:Squeeze_X}
	    & \hx_n  = e^{A(n-n_0)} \htx_n
	    \\ \label{eq:Squeeze_P}
	    & \hp_n  = e^{-A(n-n_0)} \htp_n
	\end{align}
	with $n_0$ an arbitrary real number, we have
	\begin{align}
	    \hat{H}_B
	    &
	    =
	    J
	    \sum_{n=1}^{N-1}
	    \left(
	    - \htx_{n+1} \htp_j + \htp_{n+1} \htx_j
	    \right)
	    \\
	    &
	    =
	    iJ
	    \sum_{n=1}^{N-1}
	    \left(
	    \hta_{n+1}^\dagger\hta_n
	    -
	    h.c.
	    \right)
	\end{align} 
	with $\hta_{n} = (\htx_n+i\htp_n)/\sqrt{2}$ a transformed canonical annihilation opreator. The parameter $n_0$ does not enter the Hamiltonian since $\hH_B$ is invariant under a uniform local squeezing operation that doesn't mix quadratures $\hx_n \to e^{-An_0} \hx_n, \hp_n \to e^{An_0} \hp_n $. In this section, it will be convenient to set $n_0 = 1$, so that the annihilation operators on the first stie remain unchanged $\hta_1 = \ha_1$.
	
	 The Heisenberg-Langevin equations of motion in this new basis read
	\begin{align}
	\dot{\htx}_n & 
	= 
    -i[\htx_n, \hat{H}_B]
    -
    \delta_{n1}
    \left(
    \frac{\kappa}{2}\htx_n
    +
    \sqrt{\kappa}
    \left(
    \sqrt{2}\beta + \hX^{\rm(in)}
    \right)
    \right),
    \\
	\dot{\htp}_n & = 
    -i[\htp_n, \hat{H}_B]
    -
    \delta_{n1}
    \left(
    \frac{\kappa}{2}\htp_n
    +
    \sqrt{\kappa}
    \hP^{\rm(in)}
    \right).
	\end{align}
	As expected, the response properties of $\htx_n$ and $\htp_n$ are then determined by a completely Hermitian matrix (other than the waveguide-induced decay on the first site).
	
	Using the squeezing transformations Eqs.~(\ref{eq:Squeeze_X})-(\ref{eq:Squeeze_P}) and the fact that the dynamics of the $\hx$ and $\hp$ quadratures are uncoupled, the relevant quadrature-quadrature susceptibilities read
	\begin{align}\label{eq:chi_xx}
	    &
	    \chi^{xx}(n,m;t)
	    =
	    -i\langle[\hx_n(t), \hp_m(0)]\rangle
	    =
	    e^{A(n-m)}
	    \tilde{\chi}^{xx}(n,m;t)
	    \\ \label{eq:chi_pp}
	   &
	   \chi^{pp}(n,m;t)
	    =
	    i\langle [\hp_n(t), \hx_m(0)] \rangle
	    =
	    e^{-A(n-m)}
	    \tilde{\chi}^{pp}(n,m;t)
	   \\\label{eq:chi_xp}
	   &
	   \chi^{xp}(n,m;t)
	    =
	    i\langle [\hx_n(t), \hx_m(0)] \rangle
	    =
	   0
	    \\ \label{eq:chi_px}
	   &
	   \chi^{px}(n,m;t)
	    =
	    -i\langle [\hp_n(t), \hp_m(0)] \rangle
	    =
	   0
	\end{align}
	where $\tilde{\chi}^{\alpha \beta}(n,m;t)$ are quadrature response functions of a regular (i.e. reciprocal particle-conserving)  tight-binding chain with a waveguide attached to the first site. Note that our convention differs from that used in the condensed matter community, where $\chi^{\alpha \beta}(n,m;t)$ is the response of quadrature $\alpha$ to a force which couples to $\beta$ in the Hamiltonian.  Computing the quadrature-quadrature susceptibilities $\chi^{\alpha \beta}(n,m;t)$ of our non-reciprocal system is then no more complicated than finding the susceptibilities of a reciprocal tight-binding chain $\tilde{\chi}^{xx }(n,m;t)$ and $\tilde{\chi}^{pp}(n,m;t)$. 
	
	The susceptibilities of the Hatano-Nelson model Eq.~(\ref{eq:Hatano-Nelson}) are computed in a similar manner. There, instead of a local squeezing transformation, one makes a so called imaginary gauge transformation $\ket{n} \to e^{A (n-j_0)} \ket{n}$ and $\bra{n} \to e^{-A(n-j_0)} \bra{n}$. In this new gauge, the Hamiltonian is Hermitian and completely independent of $A$. The factorization of Eq.~(\ref{eq:Susceptibility}) as $\chi(n,m;t) = e^{A(n-m)}\tilde{\chi}(n,m;t)$ immediately follows.

\section{Particle-conserving susceptibilities}\label{app:chi_particle}
Although so far we've only considered quadrature-quadrature response functions, the fact that we can map our Hamiltonian onto a particle conserving one makes it so that it is much simpler to keep track of the dynamics of the single squeezed mode $\hta_n$. Indeed, we have
\begin{align} \label{eq:quad_diag}
    &\tilde{\chi}^{xx}(n,m;t)
    =
    \tilde{\chi}^{pp}(n,m;t)
     =
    \text{Re}
    \:
    \tilde{\chi}(n,m;t)
    \\ \label{eq:Suscept_Zero}
    &
    \tilde{\chi}^{px}(n,m;t)
    =
    -
    \tilde{\chi}^{xp}(n,m;t)
    =
    \text{Im}
    \: 
    \tilde{\chi}(n,m;t)
\end{align}
where
\begin{align}
    \tilde{\chi}(n,m;t)
    =
    \langle
    [\hta_n(t), \hta_m^\dagger(0)]
    \rangle.
\end{align}
Because our Hamiltonian is quadratic in boson opeators and conserves total quasiparticle number, we can readily use the single-particle formalism to find the relevant susceptibilities. If we let $\ket{n}$ denote a position eigenket, we then have
\begin{align}
    \tilde{\chi}(n,m;t)
    =
    \bra{n}
    e^{-i t
    \left(
    \boldsymbol{\tilde{H}}
    -i \frac{\boldsymbol{\kappa}}{2}
    \right)
    }
    \ket{m}
\end{align}
with
\begin{align}
    &\boldsymbol{\tilde{H}}
    =
     i J 
    \left(
    \sum_{n=1}^{N-1}
    \ket{n+1}\bra{n}
    -
    h.c.
    \right)
    \\
    & 
    \boldsymbol{\kappa}
    =
    \kappa \ket{1}\bra{1}
\end{align}

It is more convenient to write the susceptibilities in the frequency domain:
\begin{align}
    \tilde{\chi}[n,m;\omega]
    &
    =
    \int_0^\infty
    dt
    \chi(n,m;t) e^{i\omega t}
    \\
    &
    =
    \bra{n}
    \frac{i}{\omega \boldsymbol{1}
    -\boldsymbol{\tilde{H}}
    +
    i \frac{\boldsymbol{\kappa}}{2}}
    \ket{m}
\end{align}
We'll first compute the susceptibilities without the effects of $\epsilon$ or $\kappa$, that is
\begin{align}
    \tilde{\chi}_0[n,m;\omega]
    =
    \bra{n}
    \frac{i}{\omega \boldsymbol{1}
    -\boldsymbol{\tilde{H}}
    }
        \ket{m}
\end{align}
Written out explicitly, the matrix elements of the susceptibility for a finite open chain then satisfy the difference equation
\begin{align}\label{eq:Diff_Eq}
    i \tilde{\chi}_0[n-1,m;\omega]
    -
    \frac{\omega}{J} \tilde{\chi}_0[n,m;\omega]
    -
    i \tilde{\chi}_0[n+1,m;\omega]
    =
    -\frac{i \delta_{nm}}{J}
\end{align}
with boundary conditions $\tilde{\chi}_0[0, m; \omega] = \tilde{\chi}_0[N+1,m;\omega] = 0$. The exact form of the susceptibility matrix is known; here for the sake of completeness we quickly sketch how to obtain it. First, we note that Eq.(\ref{eq:Diff_Eq}) has the form of a translationally invariant Green's function problem in the index space $n$, with $-i \delta_{nm}/J$ acting as a source term. The general solution will then consist of a linear combination of the source free solution and a convolution (in the index space $n$) of the source with the homogeneous solution. 

The source free solution, which satisfies
\begin{align}
    i \tilde{\chi}^{\rm sf}_0[n-1,m;\omega]
    -
    \frac{\omega}{J} \tilde{\chi}^{\rm sf}_0[n,m;\omega]
    -
    i \tilde{\chi}_0^{\rm sf}[n+1,m;\omega]
    =
    0
\end{align}
is precisely (up to a factor of $i$) the recursion relation that defines $T_n(\omega/2J)$ and $U_n(\omega/2J)$, the Chebyshev polynomials of the first and second kind respectively. Since $U_{-1}(\omega/2J) = 0$, and given our boundary condition $\tilde{\chi}_0[0,m;\omega] = 0$, we conclude that the source free solution is 
\begin{align}
    \tilde{\chi}_0^{\rm sf}[n,m; \omega]
    =
    c_mi^n U_{n-1}(\frac{\omega}{2J})
\end{align}
with $c_m$ a constant that will be used to satisfy the second boundary condition. The full solution to Eq.~(\ref{eq:Diff_Eq}) is then
\begin{align}
    \tilde{\chi}_0[n,m\omega]
    =
    c_mi^n U_{n-1}(\frac{\omega}{2J})
    -\frac{i}{J}i^{n-m} U_{n-m-1}(\frac{\omega}{2J})\Theta(n-m)
\end{align}
with $\Theta(n-m)$ the Heaviside step function (where $\Theta(0) = 0$). Enforcing the second boundary condition $\tilde{\chi}_0[N+1,m;\omega] = 0$ yields
\begin{align}\label{eq:Bare_Suscept}
    \tilde{\chi}_0[n, m; \omega]  
    =
    i^{1+n-m}
    \frac{U_{\min(n,m)-1}(\frac{\omega}{2J}) U_{N-\max(n,m)}(\frac{\omega}{2 J})}{J U_{N}(\frac{\omega}{2J})}
\end{align}

We now turn our attention to computing the response functions in the presence of the waveguide on the first site. Formally, this introduces a local term $-\kappa/2 \delta_{n,1} \delta_{m,1}$ to the the dynamical matrix. The full susceptibilities $\tilde{\chi}[n,m;\omega]$ can then readily be solved algebraically using Dyson's equation 
\begin{align}\label{eq:Dyson}
    \tilde{\chi}[n,m;\omega]
    &
    =
    \tilde{\chi}_0[n,m;\omega]
    -
    \frac{\kappa}{2}
    \tilde{\chi}_0[n,1;\omega]
    \tilde{\chi}[1,m;\omega]
    \\ \nonumber
    &
    =
   \tilde{\chi}_0[n,m;\omega]
    -
    \frac{
    \frac{\kappa}{2}
    \tilde{\chi}_0[n,1;\omega]
    \tilde{\chi}_0[1,m;\omega]
    }
    {
    1+\frac{\kappa}{2}\tilde{\chi}_0[1,1;\omega]
    }.
\end{align}
Since there is only a driving force on the first site and we are only interested in the response on the first site, we must only compute $\tilde{\chi}[n,1;\omega]$ and $\tilde{\chi}[1,m;\omega]$:
\begin{align}\label{eq:chi_n1}
    \tilde{\chi}[n,1;\omega]
    =
    i^{n}
    \frac{U_{N-n}(\frac{\omega}{2J})}
    {J U_N(\frac{\omega}{2J})+i \frac{\kappa}{2}U_{N-1}(\frac{\omega}{2J})}
    \\ \label{eq:chi_1n}
    \tilde{\chi}[1,m;\omega]
    =
    -i^{-m}
    \frac{U_{N-m}(\frac{\omega}{2J})}
    {J U_N(\frac{\omega}{2J})+i \frac{\kappa}{2}U_{N-1}(\frac{\omega}{2J})}
\end{align}
Because $\tilde{\chi}^{px}(n,m;t) = - \tilde{\chi}^{xp}(n,m;t) = 0$, from Eq.~(\ref{eq:Suscept_Zero}) we conclude that 
 $\tilde{\chi}[n,m;\omega] = \tilde{\chi}^{xx}[n,m;\omega] = \tilde{\chi}^{pp}[n,m;\omega]$. With this result and Eqs.(\ref{eq:chi_xx})-(\ref{eq:chi_pp}), we now have all the relevant quadrature-quadrature susceptibilities. 

\section{Total photon number}\label{app:n_bar}
Let us compute the total steady-state intracavity photon number on each site to zeoreth order in $\epsilon$. To do so, we must solve the Heisenberg-Langevin equations for the cavity annihilation operators $\ha_n$. Recall that we were able to define new squeezed annihilation and creation operators 
\begin{align}
    \ha_n 
    =
    \cosh(A(n-1)) \hta_n
    +
    \sinh(A(n-1)) \hta_n^\dagger
\end{align}
where the Hamiltonian $\hH_B$ conserved the total number of quasiparticles. Thus, the total number of photons on site $n$ reads
\begin{align}\nonumber
    \langle \ha_n^\dagger \ha_n \rangle
    &
    =
    \cosh(2A(n-1)) 
    \langle \hta_n^\dagger \hta_n\rangle
    \\ \nonumber
    &
    +
    \sinh(2A(n-1))\Re(\langle \hta_n \hta_n \rangle)
    \\
    &+
    \sinh^2(A(n-1))
\end{align}
The last term is due to noise that enter the port on site 1 and is turned into real photons by the parametric amplifier-type interactions. We can readily solve the Heisenberg-Langevin equations for the squeezed modes $\hta_n$:
\begin{align} \label{eq:HL_Solved} \nonumber
    \hta_n(t)
    &
    =
    \tilde{\chi}(n,m;t)
    \hta_m(t)
    \\ \nonumber
    &
    -
    \sqrt{\kappa}\beta 
    \int_0^{t}
    dt' \tilde{\chi}[n,1;t-t']
    \\ 
    &
    -\sqrt{\kappa}
    \int_0^{t}
    dt' \tilde{\chi}[n,1;t-t'] \ha^{\rm(in)}(t')
\end{align}
where $\ha^{\rm (in)}(t) = (\hX^{\rm(in)}(t) + i \hP^{\rm (in)}(t))/\sqrt{2}$ is the operator equivalent of Gaussian white noise. Note that we're using Einstein summation notation. Assuming a zero temperature environment we have in the steady-state
\begin{align}
    \langle \hta_n^\dagger \hta_n\rangle
    =
    \langle \hta_n \hta_n\rangle
    =
    \kappa
    \beta^2
    |\tilde{\chi}[n,1; \omega =0]|^2.
\end{align}
Using Eq.~(\ref{eq:chi_n1}). we obtain
\begin{align}\label{eq:photon_number}
    \langle \ha^\dagger_n \ha_n \rangle
     &
     =
     \frac{4\beta^2}{\kappa} e^{2A(n-1)} \sin^2 \frac{\pi}{2}n
     \\ \nonumber
     &
     +
     \sinh^2(A(n-1))
\end{align}
where we've assumed (and will do so throughout) that $N$ is odd. Summing Eq.~(\ref{eq:photon_number}) over all lattice sites gives
\begin{align}\label{eq:Total_n}
    \bar{n}_{\rm tot}
    &=
\frac{4 |\beta|^2}{\kappa}
\frac{e^{2A(N+1)}-1}{e^{4A}-1}
\\ \nonumber
&
+
\frac{1}{4}
\left(
\frac{\sinh(A(2N-1))}{\sinh(A)}
-
(2N-1)
\right)
\\ \nonumber
&
= 
\bar{n}_N \frac{1-e^{-2A(N+1)}}{1-e^{-4A}}+
\bar{n}_{\rm vac}
\end{align}
with $\bar{n}_{\rm vac}$ the photons that are present due to amplified vacuum fluctuations. Thus, the ratio of the average photon number on the last site to the total number of photons $Z(A)$ is
\begin{align}
    Z(A)
    =
    \left(
    \frac{1-e^{-2A(N+1)}}{1-e^{-4A}}
    +
    \frac{\bar{n}_{\rm vac}}{\bar{n}_N}
    \right)^{-1}
\end{align}
In the limit where $|\beta|^2/\kappa$ is large, the coherent photons dominate $\bar{n}_{\rm vac}$, which we can ignore. We then have
\begin{align}
    Z(A) = \frac{1-e^{-4A}}{1-e^{-2A(N+1)}}
    =
    1-\mathcal{O}(e^{-4A})
\end{align}
as in the main text.

\section{QFI for $\hat{V}_{NHSE}$}\label{app:SNR}
We are now in a position to compute $\text{QFI}_{\tau}(N)/\bar{n}_{\rm tot}$ for any choice of perturbation $\hat{V}$. Recall that that in the large $\beta$ limit of interest, the QFI coincides with SNR squared, optimizing over the homodyne angle $\phi$ ( see Eqs.\ref{eq:QFI}). As is written in the main text, see Eq.(\ref{eq:Signal_First_Order}), the steady-state signal takes the form
\begin{align}
    \mathcal{S}_{\tau}(N,\epsilon)
    =
    \sqrt{\kappa \tau}
    |
    \Re
    [
    e^{-i \phi}
    (
    \delta \langle \hx_1 \rangle^{\rm ss}
    +
    i
    \delta \langle \hp_1 \rangle^{\rm ss}
    )
    ]
    |
\end{align}
with $\delta \langle \hx_1 \rangle^{\rm ss} $ and $\delta \langle \hp_1 \rangle^{\rm ss} $ the steady state linear response of the site-1 average quadrature amplitude to a non-zero $\epsilon$. The signal will depend on the phase of the coherent drive $\beta$ which we take to be real, as in the main text. Our conclusion that $\hat{V}_{NHSE}$ does not have an exponentially large QFI/$\bar{n}_{\rm tot}$ is independent of the phase of $\beta$, as will become evident. This choice of phase is equivalent to driving the $\hx_1$ quadrature with a force $-\sqrt{2 \kappa}\beta$, so that the signal is
\begin{align}
    \mathcal{S}_{\tau}(N,\epsilon)
    =
    \kappa \beta \sqrt{2\tau}
    |
    \Re
    [
    e^{-i \phi}
    (
    \delta \chi^{xx}[1,1;0]
    +
    i
    \delta \chi^{px}[1,1;\omega]
    )
    ]
    |
\end{align}
The form of the responses $\delta \chi^{xx}[1,1;0]$ and $\delta \chi^{px}[1,1;0]$ will depend on $\hat{V}$. For the non-local hopping perturbation
\begin{align}
    \hat{V}_{\rm NHSE}
    =
    e^{i \varphi}\ha^\dagger_1 \ha_N 
    +    
    e^{-i \varphi }\ha^\dagger_N \ha_1
\end{align}
the change to the equations of motion induced by $\epsilon$ to the quadratures on the first site read:
\begin{align}\label{eq:Couple_1}
    &\delta \dot{\hx}_1
    =
    \epsilon
    \left(
    \sin \varphi \: \hx_N
    +
    \cos \varphi \: \hp_N
    \right)
    \\ \label{eq:Couple_2}
    &
    \delta \dot{\hp}_1
    =
    \epsilon
    \left(
    -\cos \varphi \: \hx_N+\sin \varphi \: \hp_N
    \right).
\end{align}
First order perturbation theory then yields
\begin{align}
    &\delta \chi^{xx}[1,1;0]
    =
    \chi^{xx}[1,1;0]
    (\epsilon \sin \varphi)
    \chi^{xx}[N,1;0]
    \\
    &
    \delta \chi^{px}[1,1;0]
    =
    \chi^{pp}[1,1;0]
    (-\epsilon \cos \varphi)
    \chi^{xx}[N,1;0]
\end{align}
where $\chi^{\alpha \alpha}[n,m;\omega]$ the susceptibilities of the unperturbed system, which we computed in Appendix \ref{app:chi_quadrature} and Appendix \ref{app:chi_particle}. The salient feature is that $\chi^{xx}[n,m;\omega] \propto e^{A(n-m)}$ and $\chi^{pp}[n,m;\omega] \propto e^{-A(n-m)}$ due to the phase-dependent chiral propagation. With the factor of $\chi^{xx}[N,;0]$, it would appear that we have we do in fact have an exponentially large response. Yet it precisely this terms which controls the number of coherent photons on site $N$, since $\bar{n}_N = \kappa |\beta|^2|\chi^{xx}[N,1,0]|^2$. Expressing the signal in terms of  $\bar{n}_N$ gives 
\begin{align}\label{eq:Boring_SNR}
    \mathcal{S}_{\tau}(N,\epsilon)
    =
    \sqrt{8\tau \kappa \bar{n}_N}  |\frac{\epsilon}{\kappa}|
    |
    \sin (\varphi-\phi)
    |
\end{align}
where we've used $\chi^{xx}[1,1;0] = \chi^{pp}[1,1;0] = 2/\kappa$ for a chain with a odd number of sites. The form of Eq.~(\ref{eq:Boring_SNR}) makes it evident that SNR$/\sqrt{\bar{n}_{\rm tot}}$ doesn't scale exponentially with system size, and therefore neither does QFI$/\bar{n}_{\rm tot}$.

Despite the perturbation having coupled the two effective Hatano-Nelson chains with an amplitude of $\epsilon \cos \varphi $ (see Eqs.~(\ref{eq:Couple_1} and \ref{eq:Couple_2})), this is not enough to ensure a large SNR$/\sqrt{\bar{n}_{\rm tot}}$. The non-local form of $\hat{V}_{\rm NHSE}$ implies that a wavepacket only experiences unidirectional amplification before exiting the waveguide. In contrast, the perturbation $\hat{V}_N = \ha^\dagger_N \ha_N$ studied throughout the main text allows for amplification before and after interacting with $\epsilon$.

\section{Single-Pole Approximation}\label{app:Single_Pole}
As mentioned in the main text, we need to understand finite-time dynamics of our non-Hermitian lattice sensor. 
While we have the exact frequency-space susceptibilities through Eqs.(\ref{eq:chi_xx}-\ref{eq:chi_px}) and Eqs.~(\ref{eq:chi_n1})-(\ref{eq:chi_1n}) to zeroth-order in $\epsilon$, Fourier transforming to the time-domain becomes an intractable problem. Note that this is only true of the signal: to zeroeth order in $\epsilon$, the noise is always vacuum.

There is however an exact form of the SNR in the limit  where the hopping is infinite $J \to\infty$. In this limit the susceptibilities Eqs.~(\ref{eq:chi_n1})-(\ref{eq:chi_1n}) take the form
\begin{align}\label{eq:Single_Pole_N1}
    \tilde{\chi}[N,1;\omega]
    =
    \frac{2i^{N}}{N+1}
    \frac{1}
    {\omega+i \frac{\kappa}{N+1}}
    \\ \label{eq:Single_Pole_1N}
    \tilde{\chi}[1,N;\omega]
    =
    \frac{-2i^{-N}}{N+1}
    \frac{1}
    {\omega+i \frac{\kappa}{N+1}}
\end{align}
such that the width of the zero mode is $\kappa/(N+1)$. The Fourier transform of each susceptibility (and their product, which is what determines linear response) is then easily computed. 

The change to the cavity quadrature amplitude $\hp_1$ at a time $t$ in response to the perturbation $\epsilon$ can be found using Eq.~(\ref{eq:HL_Solved}) and first order perturbation theory
\begin{align} \nonumber
     \langle \hp_1(t) \rangle 
     &=
     -\sqrt{2\kappa}\beta
     \int_0^t dT \delta \chi^{px}(1,1;T)
     \\ 
     &=
     \sqrt{2 \kappa} \epsilon \beta
     \int_0^t dT
     \int_0^T dT'
     \chi^{pp}(1,N:T-T')
     \chi^{xx}(N,1;T')
\end{align}
Using $\chi^{xx}(n,m;t) = e^{A(n-m)} \tilde{\chi}(n,m;t)$, $\chi^{pp}(n,m;t) = e^{-A(n-m)} \tilde{\chi}(n,m;t)$, Eqs.~(\ref{eq:Single_Pole_N1}) and (\ref{eq:Single_Pole_1N}) we get
\begin{align}
\langle \hp_1(t) \rangle 
    =
    -\epsilon 
    \sqrt{2 \kappa}\beta(\frac{2}{N+1})^2 e^{2A(N-1)}
    \int_0^{t}dT
    T e^{-\frac{\kappa T}{N+1}}
\end{align}
From which we obtain the signal
\begin{align}
    &\mathcal{S}_{\tau}(N,\epsilon, J \rightarrow \infty)
    =
    \\ \nonumber
    &\frac{\sqrt{2} \kappa |\beta||\epsilon|}{\sqrt{\tau}}
    (\frac{2}{N+1})^2
    e^{2A(N-1)}
    \int_0^\tau dt \int_0^{t} dT
    T e^{-\frac{\kappa T}{N+1}}
\end{align}
whereas the noise is always just $\mathcal{N}_\tau(N,\epsilon) = 1/\sqrt{2}$. The SNR for finite $\tau$ is then
\begin{align} 
	    &\text{SNR}_\tau(N, \epsilon,J \rightarrow \infty)
	        =
	    \\ 
	    &
	        \sqrt{ \frac{\tau}{\tau^*_{M}(N) }}
	        \left(
	            1+e^{-\frac{\tau}{t_{esc}(N)}}
	            -\frac{2 t_{esc}(N)}{ \tau }
	            (1-e^{-\frac{\tau}{t_{esc}(N)}})
	    \right)
	\nonumber
\end{align}
	where recall
\begin{align}
    	\tau^{*}_{M}(N)
	        =
	    \frac{1}{16 Z(A) \Bar{n}_{\rm tot} \kappa}
	    \left(\frac{\kappa}{\epsilon_0} \right)^2
	    e^{-2A(N-1)}
\end{align}
is the measurement time when the steady-state expression holds.

We now want to find the measurement time $\tau_M^{J = \infty}(N)$ where in both the weak and strong measurement limit. In the weak measurement limit $\tau^{*}_{M}(N) \gg t_{esc}(N)$, we recover the steady state result $\tau_M^{J = \infty}(N) = \tau_M^*(N)$. In the strong measurement limit $ \tau_M^{*}(N) \ll t_{esc}(N)$, we seek the leading order contribution to the measurement time. To that end, let us define
\begin{align}
\gamma = \frac{\tau^{J = \infty}_M(N)}{\tau_M^*(N) }   
\end{align}
so that
\begin{align}
     \left[
     \sqrt{\gamma}
     \left(
     1+e^{-\gamma \frac{\tau^*_M(N)}{t_{esc}(N)}}
     \right)
     -
     \frac{2 t_{esc}(N)}{\tau_M^*(N) \sqrt{\gamma}}
     \left(
     1-e^{-\gamma \frac{\tau_M^*(N)}{t_{esc}(N)}}
     \right)
     \right]^2
     =
     1
\end{align}
Assuming that $\gamma \tau_M^*(N)/t_{esc}(N)$ is small (which can be verified to be self-consistent after solving for $\gamma$), then we can Taylor expand the exponential to third order and obtain
\begin{align}
    \gamma^5
    =
    \left(
    \frac{\sqrt{6}t_{esc}(N)}{ \tau_M^*(N)}
    \right)^4
\end{align}
from which 
\begin{align}
    \tau^{J=\infty}_M(N)
    =
    \sqrt{6}t_{esc}(N)\sqrt[\leftroot{-2}\uproot{2}5]{\frac{ \tau_M^*(N)}{\sqrt{6}t_{esc}(N)}}
\end{align}
 as in the main text.

\section{Non-perturbative effects of $\epsilon_0$ to total photon number and output field}\label{app:Non-Pert}
We now want to consider the full effect of $\epsilon_0$ on the output field. To do so, we must compute the susceptibilities $\chi^{\alpha \beta}_{\epsilon_0}[1,1;\omega]$ to all orders in $\epsilon_0$. The full Heisenberg-Langevin equations are
	\begin{align}\label{eq:HL_x_2}
	\dot{\hx}_n & 
	= 
    -i[\hx_n, \hat{H}_B+\epsilon_0 \ha^\dagger_N \ha_N]
    -
    \delta_{n1}
    \left(
    \frac{\kappa}{2}\hx_n
    +
    \sqrt{\kappa}
    \hX^{\rm(in)}
    \right)
    \\
    \label{eq:HL_p_2}
	\dot{\hp}_n & = 
    -i[\hp_n, \hat{H}_B+\epsilon_0 \ha^\dagger_N \ha_N]
    -
    \delta_{n1}
    \left(
    \frac{\kappa}{2}\hp_n
    +
    \sqrt{\kappa}
    \hP^{\rm(in)}
    \right),
	\end{align}
where as in the main text we've incorporated the drive tone amplitude in the definition of the input operators $\langle \hX^{\rm (in)}\rangle = \beta$ and $\langle \hP^{\rm (in)}\rangle = 0$.

Our strategy for solving the Heisenberg-Langevin equations will be nearly identical to that presented in Appendix \ref{app:chi_quadrature}. The key difference is that our squeezing transformation is now defined as
\begin{align}
    	\label{eq:Squeeze_X_2}
	    & \hx_n  = e^{A(n-N)} \htx_n
	    \\ \label{eq:Squeeze_P_2}
	    & \hp_n  = e^{-A(n-N)} \htp_n.
\end{align}
In this new frame, the Hamiltonian $\hH_B + \epsilon_0 \hta_N^\dagger \hta_N$ preserves total quasiparticle number $\hat{\tilde{N}}$. The Heisenberg-Langevin equations are then
	\begin{align}
	\dot{\htx}_n & 
	= 
    -i[\htx_n, \hat{H}_B+\epsilon_0 \hta_N^\dagger \hta_N]
    -
    \delta_{n1}
    \left(
    \frac{\kappa}{2}\htx_n
    +
    e^{A(N-1)}
    \sqrt{\kappa}
     \hX^{\rm(in)}
    \right)
    \\
	\dot{\htp}_n & = 
    -i[\htp_n, \hat{H}_B+\epsilon_0 \hta_N^\dagger \hta_N]
    -
    \delta_{n1}
    \left(
    \frac{\kappa}{2}\htp_n
    +
    e^{-A(N-1)}
    \sqrt{\kappa}
    \hP^{\rm(in)}
    \right)
	\end{align}
Crucially, we can immediately conclude that our chain is dynamically stable for any value of $\epsilon_0$ and $A$: the spectrum is determined by the particle conserving Hamiltonian $\hH_B+\epsilon_0 \hta_N^\dagger \hta_N $ and dissipation $\kappa/2$ on the first site.

Using these squeezing transformations in conjunction with Eqs.~(\ref{eq:quad_diag}) and (\ref{eq:Suscept_Zero}), we obtain the full form of the susceptibilities:
\begin{align}\label{eq:chi_xx_full}
    \chi^{xx}_{\epsilon_0}(n,m;t) 
    &= e^{A(n-m)} 
    \Re \tilde{\chi}_{\epsilon_0}(n,m;t)
    \\
    \chi^{pp}_{\epsilon_0}(n,m;t) 
    &= e^{-A(n-m)} 
    \Re \tilde{\chi}_{\epsilon_0}(n,m;t)
    \\ \label{eq:chi_xp_full_1}
    \chi^{xp}_{\epsilon_0}(n,m;t) 
    &=
    -
    e^{-A(2N-n-m)}
    \Im \tilde{\chi}_{\epsilon_0}(n,m;t)
    \\ \label{eq:chi_xp_full}
    \chi^{px}_{\epsilon_0}(n,m;t) 
    &=
    e^{A(2N-n-m)}
    \Im \tilde{\chi}_{\epsilon_0}(n,m;t)
\end{align}
where $\tilde{\chi}_{\epsilon_0}(n,m;t)$ is the susceptibility matrix of the complex modes $\hta_n$. 

 We already have the susceptibilities $\tilde{\chi}[n,m;\omega]$ of our tight-binding chain which incorporate the full effects of the the waveguide via Eq.~(\ref{eq:Dyson}). The frequency shift on the last site adds a term $-i\epsilon_0 \delta_{n,N} \delta_{m,N}$ to the dynamical matrix. Dyson's equation in frequency space the gives:
\begin{align}
    \tilde{\chi}_{\epsilon_0}[n,m;\omega]
    &=
    \tilde{\chi}[n,m;\omega]
    -
    i\epsilon_0
    \tilde{\chi}[n,N;\omega]
    \tilde{\chi}_{\epsilon_0}[N,m;\omega]
    \\
    &=
    \tilde{\chi}[n,m;\omega]
    -
    \frac{i \epsilon_0 \tilde{\chi}[n,N;\omega] \tilde{\chi}[N,m;\omega] }{1+i \epsilon_0 \tilde{\chi}[N,N;\omega]}.
\end{align}
Since there is a driving force only on the first site, we just need to find the susceptibilities to a force on the first site:
\begin{align}\label{eq:chi_full}
    \tilde{\chi}_{\epsilon_0}[n,1;\omega]
    =
    i^{n}
    \frac{U_{N-n}(\frac{\omega}{2J})-\frac{\epsilon_0}{J} U_{N-1-n}(\frac{\omega}{2J})}
    {J U_N(\frac{\omega}{2J})+(i\frac{\kappa}{2}-\epsilon_0)U_{N-1}(\frac{\omega}{2J})-i\frac{\epsilon_0}{J} \frac{\kappa}{2}U_{N-2}(\frac{\omega}{2J})}
\end{align}

We now compute the steady state total photon number $\bar{n}_{\rm tot}(\epsilon_0)$ when $\epsilon_0 \neq 0$. Recall we are interested in the regime where $\epsilon_0/\kappa \ll 1$ but $e^{A(N-1)} \epsilon_0/\kappa$ is not a priori small. The form of our susceptibilities Eqs.(\ref{eq:chi_xx_full})-(\ref{eq:chi_xp_full}) implies that $A$ doesn't effect the spectrum, but just the residue of the poles as expected from our previous discussion. A non-zero value of $\epsilon_0$ changes both the coherent drive-induced photon number, in addition to drive-independent photons generated from input vacuum fluctuations. The leading order correction to the total photon number when $\epsilon_0 \neq 0$ is therefore
\begin{align}
    \bar{n}_{\rm tot}(\epsilon_0)
    =
    \bar{n}_{\rm tot}(0)
    +
    (
    c
    \frac{\beta^2}{\kappa}
    +
    d
    )
    e^{4A(N-1)}
    (\frac{\epsilon_0}{\kappa})^2 
    +
    \mathcal{O}(
    e^{4A(N-2)}
    (\frac{\epsilon_0}{\kappa})^2
    )
\end{align}
where $c$ and $d$ are constants of order unity. Thus, Eq.~(\ref{eq:Full_SNR}) gives
\begin{align}
    Q(A, \epsilon_0) = \frac{4|\beta|^2 e^{2A(N-1)}/\kappa}
    {
    \bar{n}_{\rm tot}(0)
    +
    \frac{1}{2}
    (c \frac{\beta^2}{\kappa}+d)e^{4A(N-1)}(\frac{\epsilon_0}{\kappa})^2
    +
    \mathcal{O}(e^{4A(N-2)}(\frac{\epsilon_0}{\kappa})^2)
    }.
\end{align}
With the optimal amplification factor $A^*$ 
\begin{align}
e^{4A^*(N-1)}
=
\frac{\kappa^2}{8 \epsilon_0^2}
\end{align}
in conjunction with Eq.~(\ref{eq:Total_n}), we get
\begin{align}
    Q(A^*, \epsilon_0)
    &
    =
    \left(
    \frac{1-e^{-2A^*(N+1)}}{1-e^{-4A^*}}
    +
   \mathcal{O}(e^{-2A^*(N-1)})
    \right)^{-1}
    \\
    &
    =
    1-\mathcal{O}((\frac{8 \epsilon_0^2}{\kappa^2})^{\frac{1}{N-1}})
\end{align}
As in the main text. Note that we've taken the relevant limit $\beta^2/\kappa \gg 1$ such that we can ignore the amplified vacuum fluctuations to the total photon number

With the susceptibilities Eqs.~(\ref{eq:chi_xx_full})-\ref{eq:chi_xp_full}, we can also compute the quadrature-quadrature scattering matrix. If we first 
define
\begin{align}
    s[\omega]
    &=
    1- \kappa \tilde{\chi}_{\epsilon_0}[1,1;\omega]
    \\
    &=
    \frac
    {a[\omega]+i b[\omega]}
    {a[\omega]-ib[\omega]}
\end{align}
with
\begin{align}
    &a[\omega]
    =
    J U_N(\frac{\omega}{2J})-U_{N-1}(\frac{\omega}{2J}) \epsilon_0
    \\
    &b[\omega]
    =
    \frac{\kappa}{2}\left(\frac{\epsilon_0}{J}U_{N-2}(\frac{\omega}{2J})-U_{N-1}(\frac{\omega}{2J})\right)
\end{align}
then using the input-output boundary conditions Eq.~(\ref{eq:In_Out}) we find that the scattering matrix is 
\begin{align}
    \boldsymbol{s}[\omega]
    =
    \begin{pmatrix}
    R[\omega]
    &
    -T[\omega]e^{-2A(N-1)}
    \\
    T[\omega]e^{2A(N-1)}
    &
    R[\omega]
    \end{pmatrix}
\end{align}
with
\begin{align}
    &
    R[\omega] = \frac{1}{2}
    \left(
    s[\omega]
    +
    s^*[-\omega]
    \right)
    \\
    &
    T[\omega]
    =
    \frac{1}{2i}
    \left(
    s[\omega]
    -
    s^*[-\omega]
    \right).
\end{align}
Note that $|s[\omega]|^2 = 1$, which implies $|R[\omega]|^2+|T[\omega]|^2 = 1$. The zero-frequency component of $R(\epsilon_0)$ and $T(\epsilon_0)$ are then:
\begin{align}
    R(\epsilon_0)
    =
    -\frac{(\frac{\kappa}{2})^2-\epsilon_0^2}{(\frac{\kappa}{2})^2+\epsilon_0^2}
    \\
    T(\epsilon_0)
    =
    \frac{\kappa \epsilon_0}{(\frac{\kappa}{2})^2+\epsilon_0^2}
\end{align}
as in the main text.

\bibliography{NonHermitianSensing_Bib}

\end{document}